\documentclass{article}

\usepackage{microtype}
\usepackage{graphicx}
\usepackage{subfigure}
\usepackage{booktabs} 
\usepackage{threeparttable}
\usepackage{algorithm}
\usepackage[noend]{algpseudocode}
\graphicspath{{icml2025/}}
\usepackage{hyperref}
\usepackage{balance}

\usepackage[accepted]{icml2025}

\usepackage{amsmath}
\usepackage{amssymb}
\usepackage{mathtools}
\usepackage{amsthm}
\usepackage{setspace}
\usepackage[capitalize,noabbrev]{cleveref}

\theoremstyle{plain}
\newtheorem{theorem}{Theorem}

\theoremstyle{definition}

\theoremstyle{remark}

\usepackage[textsize=tiny]{todonotes}

\icmltitlerunning{False Discovery Rate Control via Frequentist-assisted Horseshoe}

\begin{document}

\twocolumn[
\icmltitle{False Discovery Rate Control via Frequentist-assisted Horseshoe}




\begin{icmlauthorlist}
\icmlauthor{Qiaoyu Liang}{yyy1}
\icmlauthor{Zihan Zhu}{sch1}
\icmlauthor{Ziang Fu}{sch2}
\icmlauthor{Michael Evans}{yyy1}


\end{icmlauthorlist}

\icmlaffiliation{yyy1}{Department of Statistical Sciences, University of Toronto.}
\icmlaffiliation{sch1}{School of Medicine, Case Western Reserve University.}
\icmlaffiliation{sch2}{School of Statistics, Beijing Normal University.}


\vskip 0.3in
]



\printAffiliationsAndNotice{} 

\begin{abstract}
\setstretch{0.989}
The horseshoe prior, a widely used handy alternative to the spike-and-slab prior, has proven to be an exceptional default global-local shrinkage prior in Bayesian inference and machine learning. However, designing tests with frequentist false discovery rate (FDR) control using the horseshoe prior or the general class of global-local shrinkage priors remains an open problem. In this paper, we propose a frequentist-assisted horseshoe procedure that not only resolves this long-standing FDR control issue for the high dimensional normal means testing problem but also exhibits satisfactory finite-sample FDR control under any desired nominal level for both large-scale multiple independent and correlated tests. We carry out the frequentist-assisted horseshoe procedure in an easy and intuitive way by using the minimax estimator of the global parameter of the horseshoe prior while maintaining the remaining full Bayes vanilla horseshoe structure. The results of both intensive simulations under different sparsity levels, and real-world data demonstrate that the frequentist-assisted horseshoe procedure consistently achieves robust finite-sample FDR control. Existing frequentist or Bayesian FDR control procedures can lose finite-sample FDR control in a variety of common sparse cases. Based on the intimate relationship between the minimax estimation and the level of FDR control discovered in this work, we point out potential generalizations to achieve FDR control for both more complicated models and the general global-local shrinkage prior family.
\end{abstract}

\section{Introduction}
\label{sec1}

\begin{figure}[t]
	\centering
	\includegraphics[width=0.85\linewidth]{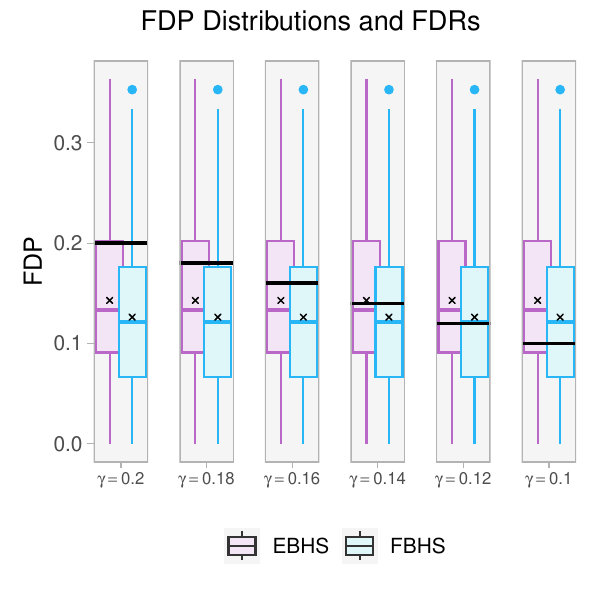}
	\vspace{-0.2cm}
	\caption{Illustration of tests using the empirical Bayes vanilla horseshoe (EBHS) and the full Bayes vanilla horseshoe (FBHS) where these tests gradually lose FDR control when the desired FDR nominal level gets tighter. Boxplots and black crosses are FDP distributions and FDRs, respectively. In this experiment,  we do 100 replications where we test 200 hypotheses in each replication under the signal proportion being 0.1.}
	\label{fig_1}
\end{figure}
In the big data era, scientists across different domains test tens of thousands of hypotheses simultaneously with the aid of modern computation power. By doing those large-scale multiple hypothesis tests, scientists make new scientific discoveries to advance science and make a better life for mankind. For example, geneticists can encounter thousands of genes in DNA microarray experiments and they need to determine whether those genes have a non-negligible association with phenotypes of interest such as disease. It is imperative that geneticists find genes that truly have significant effects on phenotypes of interest so that geneticists can do further studies in the right direction and not waste precious and limited resources. Unfortunately, many studies such as this Science paper \citep{OSC2015psy} reveal many discoveries claimed by scientists are later found to be false and not reproducible. Taking genetics as an example again, there are only a few genes that are significantly associated with phenotypes of interest despite the total number of genes being huge. In fact, the \cite{Wellcome2007Genome} has confirmed that only seven genes are found to have non-negligible effects on Type I diabetes. Thus, it is important for researchers to be cautious about proclaiming discoveries since some discoveries can be false. 

To maintain control of a suitable number of false discoveries, \cite{Benjamini1995FDR} introduce the false discovery rate (FDR) as a crucial concept and an important metric in their seminal work to serve this goal. In this work, \cite{Benjamini1995FDR} also provide a simple but powerful step-down p-value procedure to control the desired FDR. More concretely, we wish to test $m$ simultaneous null hypotheses $H_{01}, H_{02}, \ldots, H_{0 m}$. Out of the $m$ hypotheses, $m_0$ are true nulls and $m_1=m-m_0$ are true non-nulls. We can then list out all possible scenarios in a contingency table (Table \ref{table_1}). In this contingency table, T and F stand for True and False, respectively. Similarly, D and N refer to Discovery and Nondiscovery, respectively. Based on the definitions of $m_0$ and $m_1$, the sum of TN and FD is $m_0$ and the sum of FN and TD is $m_1.$ R is the number of total discoveries/rejections which is the sum of FD and TD.

\begin{table}[H]
\centering
\caption{Possible outcomes of multiple testing.}
\label{table_1}
\begin{tabular}{ccc}
\hline
                        & $H_0$ Retained & $H_0$ Rejected \\ 
$H_0$ True (Noise)      & TN             & FD             \\ 
$H_0$ False (Signal)    & FN             & TD             \\
Total                   & $m-$R          & R              \\
\hline\hline
\end{tabular}

\end{table}

The false discovery rate (FDR) is the expectation of the false discovery proportion (FDP) where the FDP is the proportion of false discoveries over total discoveries (i.e. FDP = FD/R). Using the terms associated with the classical significance test, the FDP is the proportion of false positives over the total number of tests called significant. In order to restrict the proportion of false discoveries, we would require FDR=$\mathbb{E}\text{[FDP]} \leq \gamma$ where $\gamma$ is the FDR nominal level and we define $0/0 = 0$ for the FDP. The value of $\gamma$ is typically set to be some positive number below $0.2.$ The stricter you want the control to be, the smaller the number you can set. Over the past almost thirty years, the FDR has quickly become a well-established paradigm in modern statistics and has had an enormous influence on medical research. In fact, the FDR not only transforms researchers' perspectives within the scientific community but also influences experimentation practices in the technology industry. For example, data scientists use A/B tests to test a variety of potential variations of a website to obtain business insights and make decisions on adopting the variation based on desired user metrics. 

\subsection{Global-local Shrinkage Prior}
    In Bayesian statistics and machine learning, the spike-and-slab, two-groups framework \citep{mitchell1988BVSLM,edgeorge1993VSGibbs} is classical and popular for handling sparsity. One can, however, face prohibitive computational expenses using it in high-dimensional problems. This issue motivates researchers to develop and study a new class of smooth shrinkage priors called global-local shrinkage priors \citep{polson2010spBayR} which provide a huge computational advantage over spike-and-slab two-groups structure \citep{Tadesse2021HandbookOB}. For global-local shrinkage priors, the global term provides substantial shrinkage towards zero and local terms have heavy tails so that signals would not be shrunk too much. The global parameter is routinely estimated in an empirical Bayes or a fully Bayes fashion. 
    
    The global-local shrinkage priors form a general class. Most of the existing shrinkage priors in the literature belong to the family of global-local shrinkage priors \citep{Tadesse2021HandbookOB}. Among them, the horseshoe prior is arguably most commonly employed. Consider the sparse normal means model $\left(y_j \mid \beta_j\right) \sim N\left(\beta_j, \sigma^2\right)$ for $j=1, \ldots, m$ where $\sigma^2$ is set to be $1$ without loss of generality and it is believed that many of the $\beta_j$'s equal 0. The global-local normal scale mixture shrinkage priors for sparsity introduced in \cite{Carvalho2010TheHE} can be expressed as: 
\begin{equation}
\beta_j \sim N\left(0, \xi^2\eta_j^2 \right); \ \eta_j \sim \pi(\eta_j); \ \xi \sim \pi(\xi).
\label{1hs}
\end{equation}
The well-known horseshoe prior on $\boldsymbol{\beta}$ is induced when $\pi\left(\eta_j\right) \propto\left(1+\eta_j^2\right)^{-1}$ in the global-local shrinkage structure (\ref{1hs}). 

    The global-local shrinkage priors have amazing theoretical optimal or near-optimal properties \citep{Tadesse2021HandbookOB}. In particular, the horseshoe prior has been widely adopted in Bayesian statistics, machine learning, and deep learning (\citeauthor{Tadesse2021HandbookOB}, \citeyear{Tadesse2021HandbookOB}; \citeauthor{Bhadra2019LassoMH}, \citeyear{Bhadra2019LassoMH}; \citeauthor{Bhadra2020HSDM}, \citeyear{Bhadra2020HSDM}). Designing tests, however, with frequentist FDR control using the global-local shrinkage prior, remains an open problem even in the normal means setting \citep{Bai2019NorBeta}. Figure \ref{fig_1} shows FDP distributions (i.e. boxplots) and FDRs (i.e. black crosses) when one deploys the vanilla horseshoe \citep{Carvalho2010TheHE} in the sparse normal means setting with the global parameter estimated by an empirical Bayes (EBHS) or a fully Bayes approach (FBHS). We found that both EBHS and FBHS can control FDR when the nominal levels of FDR are set to be between 0.14 and 0.2 in the sparse normal means setting. This shows the vanilla horseshoe is indeed a well-designed default prior which has the ability to control the FDR in many cases. However, this also demonstrates that the vanilla horseshoe cannot control FDR under arbitrarily chosen FDR nominal levels. 

\subsection{Two-groups Mixture Structure}
The spike-and-slab two-groups structure, however, can be used to design tests with a guarantee of any level of frequentist FDR control in the asymptotic sense \citep{Efron2008ME2GM,Sun2007OracleAA,Castillo2020OnSA}. Generally speaking, they formulate the multiple testing problem using a random two-groups mixture model: $ \beta_1, \cdots, \beta_m \stackrel{\text { i.i.d. }}{\sim} \operatorname{Bernoulli}(\tau)$ and $ Y_j \mid \beta_j \stackrel{\text { ind. }}{\sim}\left(1-\beta_j\right) F_0+\beta_j F_1 \text {. }$ The density of $F_0$ and $F_1$ are denoted by $f_0$ and $f_1$ respectively. The marginal pdf is
$
f=(1-\tau) f_0+\tau f_1 
$ where $f_0$ and $f_1$ are used for modeling the null distribution and the non-null distribution respectively. They then rely on the core test statistic, namely, the local false discovery rate (locfdr) 
\begin{equation}
\operatorname{locfdr}(z)=\frac{(1-\tau) f_0(z)}{f(z)},
\label{2locfdr}
\end{equation} 
given the $z$-score, to develop an empirical Bayes type of multiple testing procedure to separate the non-null cases from the null. \cite{Efron2001EmpiricalBA} interpret the local false discovery rate (locfdr) as the a posteriori probability of being in the null group given the $z$-score. In practice, $\tau, f_0$ and $f_1$ are unknown, but estimable from the $z$-score. Estimating $f_0$ and $f_1$ is a deconvolution problem \citep{Efron2016ebde} which is discussed intensively in the empirical Bayes literature. In this paper, we mainly focus on the development of false discovery rate control using the popular horseshoe prior (called frequentist-assisted horseshoe procedure). In addition, we point out possible strategies and generalizations to more complicated models using the general class of global-local shrinkage priors whenever appropriate. 

\textbf{Our contributions.} We make the following three
contributions in this work.

    1. \textit{The first false discovery rate control procedure using the horseshoe prior (called the frequentist-assisted horseshoe procedure) which has an extremely simple, and practical implementation.} To the best of our knowledge, the frequentist-assisted horseshoe procedure (FAHS) we propose is the first false discovery rate control procedure using the horseshoe prior with a strong theoretical guarantee. Both vanilla horseshoe benchmarks generally can not control the false discovery rate control. There is some literature (e.g. \cite{Bai2019NorBeta}) using certain global-local shrinkage priors to control FDR but these are usually being used only for a limited illustrative purpose which is far from a comprehensive study. These studies also usually illustrate FDR control when the signals are fairly dense or the FDR nominal level is set to be fairly high.
    
    It is worth pointing out that we are by no means proposing an optimal FDR control test in the sense that we control FDR exactly on its nominal level to ensure the power of this test reaches its potential maximum. Instead, our main target is to first provide a practical route to utilize handy horseshoe prior for achieving principled FDR control. The empirical Bayes approach we developed for horseshoe here can be generally applicable to other global-local shrinkage priors since they share many common characteristics. 
    \vspace{2pt}

    Notice that horseshoe prior is a default prior which is designed to be suitable in most general cases but it is not a cure-all.  Prior-data conflict \citep{evans2006pdc} can happen when the horseshoe assigns the majority of its mass to parameter values that are incompatible with the observed data. This can result in high FDPs. We shed light on how to avoid prior-data conflict and improve the potential power of FAHS by using prior-data conflict detection to adjust hyperparameters. With the Bayesian attribute in our FAHS, we can always elicit and incorporate useful domain knowledge \citep{O'Hagan2006} to alleviate the potential power loss issues whenever such information is available in certain scientific contexts. In this paper, with the usual preference, we would mainly focus on the control of FDR first.
    \vspace{2pt}
    
    In our work, we provide a comprehensive study of FDR control using our frequentist-assisted horseshoe procedure ranging from super-sparse signal cases to fairly dense signal cases. Our frequentist-assisted horseshoe procedure consistently controls FDR ranging from a relatively high FDR nominal level (e.g. 0.2) to a low FDR nominal level (e.g. 0.1) as clearly displayed in Figures \ref{fig_ind}, \ref{fig_corr}, \ref{fig_ind_full}, \ref{fig_corr_0.1}, \ref{fig_corr_0.2}, and \ref{fig_corr_0.3}. Our FAHS procedures fill the gap that there is no existing approach for the global-local shrinkage priors to control FDR with the finite sample. 
    \vspace{2pt}
    
    Motivated by a similar spirit of protecting against worst cases and by the close link between minimax estimation and FDR control \citep{song2023FDRMm}, we focus on studying minimax results of the horseshoe prior to develop our frequentist-assisted horseshoe procedure (FAHS) procedure. \cite{VanderPas2014HSconcen} provide some solid minimax asymptotic results for the horseshoe using modern minimax techniques. Using a different perspective, \cite{Piironen2017SparsityIA} define and characterize the finite-sample effective number of nonzero components $m_{\mathrm{eff}}$ of $\boldsymbol{\beta}$ for the horseshoe and some other shrinkage priors. 
    \vspace{2pt}
    
    We develop two finite-sample FDR-guaranteed frequentist-assisted horseshoe procedures (namely m-FAHS and e-FAHS procedures) by applying the results in \cite{VanderPas2014HSconcen} and \cite{Piironen2017SparsityIA} to estimate the global parameter of the horseshoe prior (i.e. $\xi$ in (\ref{1hs})) with the assistance of frequentist FDR control procedures (e.g. the BH procedure is applied in our work) while maintaining the remaining FBHS structure. 
    \vspace{2pt}
    
    Using the empirical Bayes idea, the frequentist-assisted horseshoe has the amazing feature of being Bayesian, as it can use the complete posterior distribution and provide accurate uncertainty quantification \citep{VanderPas2017AdaptivePC}, with a negligible additional computation cost. Interestingly, in later intensive simulations, the variability of FDP distributions (e.g. boxplots in Figure \ref{fig_1}) when using frequentist-assisted horseshoe procedures, is almost always smaller than the variation of FDP distributions when applying the BH procedure. 
\vspace{3.3pt}

2. \textit{Both m-FAHS and e-FAHS procedures have arbitrary finite-sample FDR control ranging from super sparse cases to relatively dense cases.} Based on our theoretical guarantee and intensive simulations, m-FAHS and e-FAHS procedures consistently control arbitrary finite-sample FDR levels by using the minimax estimator of the global parameter of the horseshoe prior \citep{VanderPas2014HSconcen}. Global parameters $\xi$'s in m-FAHS and e-FAHS even give the sharply minimax Bayesian contraction rate based on the result in \cite{song2020BSSM} which can be considered a refinement of \cite{VanderPas2014HSconcen}. We compare our m-FAHS and e-FAHS procedures with three popular FDR control frameworks specifically the BH procedure (\citeyear{Benjamini1995FDR}), the \cite{Storey2002DirFDR,Storey2003ThePF} q-value approach, and the spike-and-slab type two-groups mixture empirical Bayes (EB) approach popularized by \cite{Efron2001EmpiricalBA} and \cite{Efron2008ME2GM}. Theoretically speaking, the BH procedure has the finite sample FDR control guarantee while the q-value approach and two-groups EB approach only have the guarantee of the asymptotic FDR control. Among simulations on multiple independent and correlated tests, we observe m-FAHS, e-FAHS procedures, and the BH procedure are always able to control FDR as promised in the theory, whereas both the q-value approach and the two-groups EB approach can fail to control FDR in relatively dense signal cases not to mention super sparse signal cases. 
\vspace{3.3pt}

    We observe that the two-groups EB approach fails to provide FDR control even if the signals are relatively dense. Intuitively speaking, implementing the two-groups EB approach requires estimating $\tau, f_0$ and $f_1$ (with $f_0$ and $f_1$ being two densities) which are quite burdensome especially when the signals are sparse. Later simulations also verify our conjecture here. In comparison, m-FAHS and e-FAHS procedures, as the counterparts of the two-groups EB approach, only require the estimation of one global parameter $\xi$ while retaining the rest of the robust vanilla horseshoe structure. It is self-evident that m-FAHS and e-FAHS procedures have a much bigger chance of controlling FDR than the two-groups EB approach in scenarios where signals are so few. Indeed, intensive experiments inform us that m-FAHS and e-FAHS procedures do control FDR even when signals are super sparse. 
    \vspace{3.3pt}

3. \textit{Both m-FAHS and e-FAHS procedures are robust under dependency and model misspecification.} In the high dimensional independent and correlated normal means testing settings, m-FAHS and e-FAHS procedures consistently control FDR for any sparse situation and any FDR nominal level. Compared with BH, q-value, two-groups EB, and vanilla horseshoe procedures, we find that m-FAHS and e-FAHS procedures almost always produce FDP distributions with the smallest variability in simulations. This demonstrates the superior robustness of m-FAHS and e-FAHS procedures under dependency and model misspecification.

\textbf{Outline.} In Section \ref{sec2}, we describe the problem and
briefly review the BH, the q-value, the two-groups EB, and the vanilla horseshoe FDR control procedures. We introduce the frequentist-assisted
horseshoe (i.e. m-FAHS and e-FAHS) and give detailed illustrations in Section \ref{sec3}. In Section \ref{sec4}, we provide thorough simulations comparing all FDR control
procedures mentioned before, and we apply our m-FAHS and e-FAHS procedures to a prostate cancer
study in Section \ref{sec5}. In Section \ref{sec6}, we summarize our
findings and discuss future directions. We make a brief comparison between the two-groups model and the horseshoe-type one-group model in Appendix \ref{secA}. More background and theoretical optimal properties associated with horseshoe are also provided in Appendix \ref{secA}. In Appendix \ref{secB}, we give additional remarks about FDR control procedures in section \ref{sec2.2}. Related technical explanations about Theorem \ref{theo1} are given in Appendix \ref{secC}. In Appendix \ref{secD}, we give additional experiments to motivate the study of the global parameter of the horseshoe. In Appendix \ref{secE}, we provide comprehensive plots and discussions for FDR control procedures in complement to Section \ref{sec4} and Section \ref{sec5}. In Appendix \ref{secF}, we discuss the potential extensions of our current work in more detail.

\section{The Problem and Related FDR Control Procedures}
\label{sec2}

\subsection{The Problem}

Suppose we observe a high-dimensional $m$-component random vector $\left(y_1, \ldots, y_m\right) \in \mathbb{R}^m$, such that

\vspace{-3pt}
\begin{equation}
\label{3normalmean}
y_j \sim N\left(\beta_j, 1\right), \ j=1, \ldots, m.
\end{equation}

\vspace{-7pt}
This general framework is the basis of many high-dimensional problems including genetics studies. One can think of $y_j$'s being the observed log-fold-change values from a microarray and $\beta_j$'s being mean differential expression levels for each of thousands of genes. Under model (\ref{3normalmean}), the main goal is we want to identify the few signals (i.e. $\beta_j \neq 0$) with desired FDR control. Equivalently, we need to perform $m$ simultaneous tests, \begin{equation}
H_{0 j}: \beta_j=0 \ \text{vs.} \ H_{1 j}: \beta_j \neq 0 , \ \text{for} \ j=1, \ldots, m.
\label{4hypo}
\end{equation}

\subsection{Related FDR Control Procedures}
\label{sec2.2}
We consider the following FDR control procedures in our discussions and simulations.

\textbf{The Benjamini and Hochberg Procedure.} The Benjamini and Hochberg (BH) procedure is a data-adaptive procedure that provides a finite-sample FDR control under the desired FDR nominal level $\gamma$ for independent tests and weakly-dependent (e.g. PRDS condition in \cite{Benjamini2001BY}) tests. The algorithm to implement the BH procedure can be described below (i.e. Algorithm 1).

\begin{algorithm}
\caption{The BH algorithm}
\begin{algorithmic}[1]

\State For $\left\{{H}_{0 j}\right\}_{j=1}^m$, we construct test statistics $\left\{T_j\right\}_{j=1}^m$ and compute corresponding p-values $\left\{p_j\right\}_{j=1}^m$.

\State Rank $m$ p-values from smallest to largest (from most to least significant),
$
p_{(1)} \leq \ldots \leq p_{(m)} \text {. }
$

\State Set the threshold to be, $$k^*=\max _{1 \leq k \leq m}\left\{k: p_{(k)} \leq \frac{\gamma}{m} k\right\}.$$

\State Reject the null hypotheses $H_{01}, \cdots H_{0j}$ for $j = 1, \cdots m$ if $p_j \leq p_{\left(k^*\right)}.$

\end{algorithmic}
\end{algorithm}

\textbf{The q-value Procedure.}

The Storey's q-value \citep{Storey2003ThePF} procedure is a Bayesian analog to the
p-value which can be defined from the rejection regions of the frequentist tests. More concretely, the q-value of a test statistic $T_j$ is defined to be the smallest FDR over all rejection regions that reject $T_j$. Thus, determining all tests
with q-values $\leq \gamma$ significant is mathematically equivalent to control FDR at level $\gamma$. Recall that in Table \ref{table_1}, the sum of TN and FD is $m_0$. We then define $\pi_0 = m_0 / m.$ Using the familiar concept p-value, we then can write the estimation of q-values in terms of p-values mathematically as $$\hat{q}_{(j)}=\min \left\{\frac{\hat{\pi}_0 m p_{(j)}}{j}, \hat{q}_{(j+1)}\right\},$$ and $\hat{q}_{(m)}=\hat{\pi}_0 p_{(m)}$ where $\hat{\pi}_0$ is an estimate of $\pi_0.$ We use the default function in \textit{qvalue} package to carry out the q-value procedure throughout this paper.

\textbf{Two-groups Empirical Bayes Procedure.}

Based on the key test statistic locfdr (\ref{2locfdr}), we let $\widehat{\operatorname{locfdr}}\left(z_j\right)=(1-\hat{\tau}) \hat{f}_0\left(z_j\right)/ {\hat{f}\left(z_j\right)}$ be the estimate of locfdr $\operatorname{locfdr}\left(z_j\right)=(1-\tau) f_0\left(z_j\right)/f\left(z_j\right)$. Then, the general algorithm to perform the two-groups EB procedure can be expressed as below (i.e. Algorithm 2). \cite{Sun2007OracleAA} proved the two-groups EB procedure controls the FDR at the desired level $\gamma$ asymptotically. We use the default function in \textit{locfdr} package throughout this paper.

\begin{algorithm}
\caption{The two-groups empirical Bayes algorithm}
\begin{algorithmic}[1]

\State Let $\widehat{\operatorname{locfdr}_{(1)}}, \cdots, \widehat{\operatorname{locfdr}}_{(m)}$ be the ranked locfdr values.

\State Let $k=\max \left\{j: \frac{1}{j} \sum_{l=1}^j \widehat{\operatorname{locfdr}}_{(l)} \leq \gamma \right\}$.

\State Reject all $H_{0j}, j=1, \cdots, k$.

\end{algorithmic}
\end{algorithm}

As previously mentioned, the estimations of $f_0$ and $f_1$ (i.e. two density estimations) are called the deconvolution problem which is a fruitful topic but notoriously hard in general. There are many proposals for the deconvolution problem \citep{Efron2008ME2GM,Efron2016ebde,Sun2007OracleAA}, but the majority of them are either unstable to use because of modeling strategies and overly complicated assumptions or, involve some form of cross validation which can be computationally intensive or misleading depending on the number of observations. Thus, our FAHS procedures can serve as strong alternatives to the two-groups EB procedure.

\textbf{Vanilla Horseshoes.} Based on the global-local shrinkage prior structure in (\ref{1hs}) where $\pi\left(\eta_j\right) \propto\left(1+\eta_j^2\right)^{-1},$ we either put a half-Cauchy(0,1) hyperprior on the global parameter of the vanilla horseshoe or estimate the global parameter using data through maximum marginal likelihood estimator (MMLE). The former is called a full Bayes approach (FBHS) and the latter is called a empirical Bayes approach (EBHS). There is no theoretical FDR control guarantee for both vanilla horseshoes. We use them here as benchmarks for our frequentist-assisted horseshoes (FAHS). Let $\hat{\mu}_j$ be the posterior mean estimator. Under model (\ref{3normalmean}), if $\left|\hat{\beta}_j\right|>\frac{1}{2}\left|y_j\right|$ \citep{Carvalho2010TheHE}, then we consider $\beta_j$ as a signal (non-zero), and noise (zero) otherwise. Any operation related to posterior can be carried out using \textit{horseshoe} package and we implement the vanilla horseshoes by \textit{horseshoe} package throughout the entire paper.

\section{The Frequentist-assisted Horseshoe}
\label{sec3}

\begin{figure*}[ht] 
	\centering
	\includegraphics[width=\textwidth]{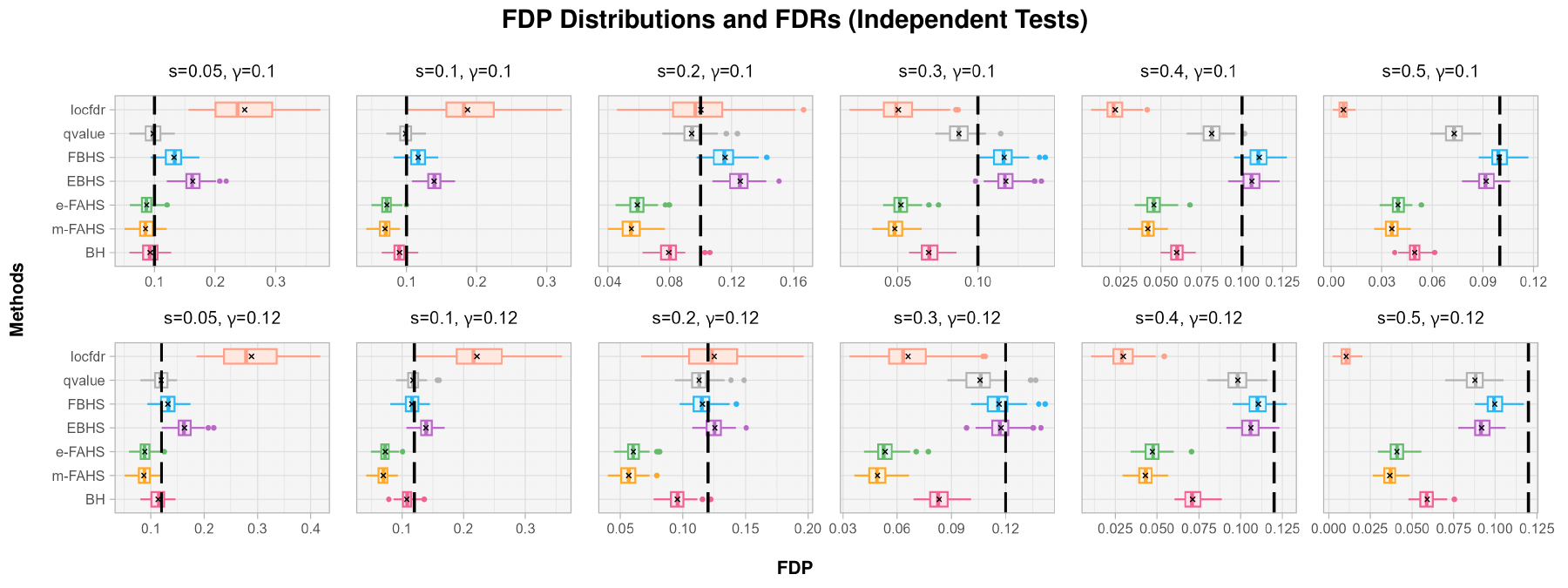} 
	\caption{ FDP distributions (boxplots) and FDRs (black crosses) for multiple FDR control methods in the multiple independent testing setting where m-FAHS and e-FAHS show more robust FDR and FDP controls than any other procedures. Black long dashed lines refer to nominal FDR levels. In this experiment, we do 100 replications where we test 10000 independent hypotheses simultaneously in each
		replication under the signal proportion ranging from 0.05 to 0.5 and the nominal FDR levels are chosen to be 0.1 and 0.12. A comprehensive figure for the multiple independent testing setting can be found in Figure 9.}
	\label{fig_ind}
\end{figure*}
In \cite{VanderPas2014HSconcen}, they prove the asymptotically optimal selection of $\xi$ for model (\ref{3normalmean}) is $\xi^*=m_1/m$ in terms of mean squared error and posterior contraction rates \citep{Ghosal2017NBI} in comparison to the true $\boldsymbol{\beta}^*.$ The global parameter of m-FAHS is chosen in this manner.

\cite{Piironen2017SparsityIA} find the finite sample optimal choice of $\xi$ in terms of the effective number of nonzero coefficients in the model (\ref{3normalmean}) is
$$\xi^{**}=\frac{m_1^{*}}{m-m_1^{*}},$$ where $m_1^{*}$ is the prior guess of $m_1$. The choice of the global parameter of e-FAHS is based on this approach. Comparing \cite{VanderPas2014HSconcen} with \cite{Piironen2017SparsityIA}, we would expect \cite{VanderPas2014HSconcen} to exhibit stronger FDR control than \cite{Piironen2017SparsityIA} since they consider a more extreme case.

The unknown quantity for both approaches is $m_1$. If we do have reliable prior knowledge of $m_1$ in certain contexts, then a reasonable choice of $\xi^*$ can then be obtained. But if we do not have this sort of prior knowledge, we have to estimate $m_1$ but there is no method proposed in the literature. Notice that we can observe the number of total discoveries (R) except $m$ by applying the BH procedure which is adaptive in terms of both $\gamma$ and $s$. By the constraint of FDR control, R and $m_1$ have a large overlapping part on the number of true discoveries (i.e. TD). Thus, R is a reasonable plug-in estimator of $m_1$. Besides, to obtain $\xi^*$ and $\xi^{**}$, we also need to adjust the estimator of $m_1$ with $m$ or $m-m_1$ where $m$ is typically quite large (e.g. easily $>10000$ in a genetic study). Thus, R/$m$ and R/($m$-R) would be reasonably close to $m_1/m$ and $m_1/(m-m_1)$ respectively. In Section \ref{sec4} and Section \ref{sec5}, We would assume we do not have prior knowledge of $m_1$ and take this plug-in approach. 

\begin{figure*}[ht] 
	\centering
	\includegraphics[width=\textwidth]{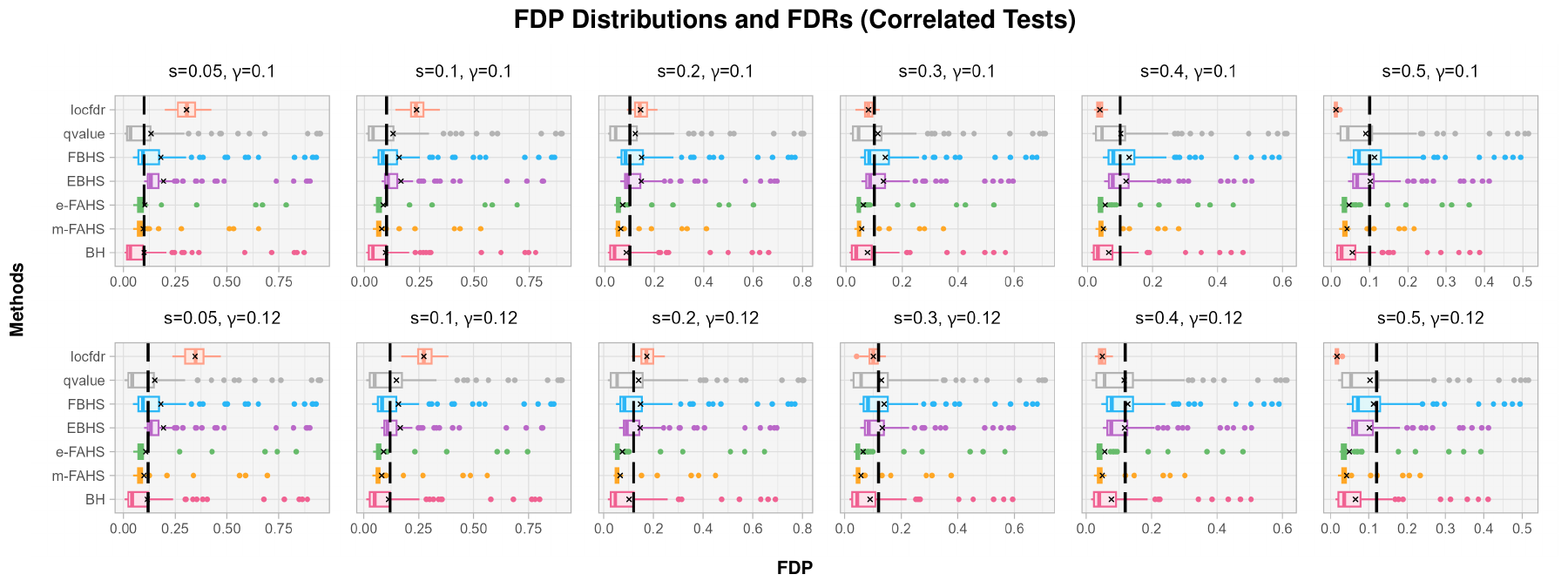} 
	\caption{FDP distributions (boxplots) and FDRs (black crosses) for multiple FDR control methods in the multiple correlated testing
		setting (equicorrelation structure with equal correlation 0.3) where m-FAHS and e-FAHS show more robust FDR and FDP controls than
		any other procedures. Black vertical dashed lines refer to nominal FDR levels. In this experiment, we do 100 replications where we test
		10000 correlated hypotheses simultaneously in each replication under the signal proportion ranging from 0.05 to 0.5 and the nominal FDR
		levels are chosen to be 0.1 and 0.12. Comprehensive figures for the multiple correlated testing setting can be found from Figure 10 to Figure 12.}
	\label{fig_corr}
\end{figure*}

Since one only needs to modify the global parameter $\xi$ provided by the vanilla horseshoe to the global parameter $\xi$ estimated as above, the implementations of both m-FAHS and e-FAHS procedures are extremely simple and straightforward. The \textit{horseshoe} package can still take care of our implementations of m-FAHS and e-FAHS procedures. The algorithm to implement the FAHS procedures can be simply summarized below (i.e. Algorithm 3).

\begin{algorithm}
\caption{The FAHS algorithms}
\begin{algorithmic}[1]

\State Obtain the number of rejections as an estimate of the number of signals (i.e. $\widehat{m_1}$) by applying the BH procedure to $\left\{H_{0 j}: \beta_j=0, j=1, \cdots, m\right\}$ at level $\gamma$.

\State Construct the frequentist-assisted estimator for $\xi$.
In m-FAHS, $\hat{\xi}=\frac{\widehat{m_1}}{m} ;$ in e-FAHS, $\hat{\xi}=\frac{\widehat{m_1}}{m-\widehat{m_1}}.$

\State Run Gibbs sampler for the normal means model with the horseshoe prior given $\hat{\xi}$, and obtain the posterior mean estimator $\hat{\beta}_j, j=1, \cdots, m$.

\State Reject the null hypotheses $H_{0 j}$ if $\left|\hat{\beta}_j\right|>\left|y_j\right| / 2$ for $j=1, \cdots m.$

\end{algorithmic}
\end{algorithm}

The $L_2$/$L_1$ posterior contraction rate $r_m$ satisfies
$$
\lim _{m \rightarrow \infty} \mathbb{E}^*\left[\pi\left(\left\|\boldsymbol{\beta}-\boldsymbol{\beta}^*\right\| \geq r_m \mid \boldsymbol{y}\right)\right]=0, \text { for any } \boldsymbol{\beta}^*,
$$
where $\mathbb{E}^*$ denotes the expectation with respect to the data generation measure of $m$ dimensional data $\boldsymbol{y}$ under true parameter $\boldsymbol{\beta}^*$. Let $ \text{log} (m / m_1)$ be $M$. Based on quantities in Table \ref{table_1}, we state a further refinement of posterior contraction rates for the horseshoe \citep{song2020BSSM} which is stronger than the result in \cite{VanderPas2014HSconcen}. This theorem is useful for verifying whether $\xi$'s in Section \ref{sec4} satisfy stricter requirements and let m-FAHS and e-FAHS enjoy sharp Bayesian contraction rates. More explanations about Theorem \ref{theo1} can be found in Appendix \ref{secC}.

\begin{theorem}(Theorem 2.1. \citep{song2020BSSM})
Given a positive constant $\omega$ and some $\alpha > 1$, if $\xi^{\alpha-1} \geq(m_1 / m)^c\{\log (m / m_1)\}^{1 / 2}$ for some $c \in(0,1+\omega / 2)$, and $\xi^{\alpha-1} \prec\{(m_1 / m) \log (m / m_1)\}^\alpha$, then
$$
\lim _{m \rightarrow \infty} \mathbb{E}^*\left(\pi\left[\left\|\boldsymbol{\beta}-\boldsymbol{\beta}^*\right\|_2 \geq C_1(\omega) (m_1 M)^{\frac{1} {2}} \mid \boldsymbol{y} \right]\right)=0,
$$

where $C_1(\omega)=\sqrt{2+\omega}+\sqrt{\omega}$ and it satisfies $\lim _{\omega \downarrow 0} C_1(\omega)=\sqrt{2}$. If furthermore, $\xi^{\alpha-1} \prec(m_1 / m)^\alpha\{\log (m / m_1)\}^{(\alpha+1) / 2}$, then
$$
\lim _{m \rightarrow \infty} \mathbb{E}^*\left(\pi\left[\left\|\boldsymbol{\beta}-\boldsymbol{\beta}^*\right\|_1 \geq m_1 C_2(\omega) {M}^{\frac{1} {2}} \mid \boldsymbol{y}\right]\right)=0,
$$
where $C_2(\omega)=\sqrt{2+\omega}+\sqrt{w^2 / 5}+\sqrt{\omega / 5}$, and it satisfies $\lim _{\omega \downarrow 0} C_2(\omega)=\sqrt{2}$.
\label{theo1}
\end{theorem}

\section{Simulation Studies}
\label{sec4}

\subsection{Multiple Independent Tests}
\label{sec4.1}
We use the following true data-generating process,
$$
y_j = \beta_j^{}+\epsilon_j, \quad j=1,2, \cdots, m,
$$
where $\beta_j \sim s N\left(0, \psi^2\right)+(1-s) \delta_0$ and $\epsilon_j \sim N(0,1)$.

Let $ m=10000, \ \psi=5,$ and $s$ are chosen to be $0.05, 0.1,0.2,0.3,0.4,0.5.$ with $s=0.05$ being the most sparse signal case. Let $\gamma$ be chosen to be $0.1, 0.12, 0.14, 0.16, 0.18, 0.2.$ with $\gamma=0.1$ being the strictest case. In the experiment, we do 100 replications for every setting (i.e. any combination of $s$ and $\gamma$) where we use multiple FDR control methods to test 10000 independent hypotheses simultaneously in each replication. The signal proportion $s$ ranges from 0.05 to 0.5 and the nominal FDR level $\gamma$ can be chosen between 0.1 and 0.2. The entire result is in Appendix \ref{secE}. In Figure \ref{fig_ind}, we show the FDP distributions and FDRs when the FDR nominal level is chosen to be 0.1 or 0.12. 

We observe two-groups EB procedure loses FDR control in almost all sparse signal settings ($s = 0.05, 0.1, \text{and} \ 0.2$) in terms of the entire result. Even when the signal becomes dense enough, the variability of FDP distributions for this procedure is much larger than any other procedure. Until half of the observations are signals (i.e. most dense signal case in our experiment), the variability of FDP distributions for it becomes comparable with the variability of FDP distributions for other procedures. Regardless of the situation, our FAHS procedures, as the alternative of the two-groups EB procedure, almost always produce the strongest FDR control and smallest variation of FDP distributions among all methods. Similar to what we observed in Figure \ref{fig_1}, both default horseshoe procedures (EBHS and FBHS) lose FDR control as the FDR nominal level tightens. Although the q-value procedure controls FDR in most experiments, it can easily lose FDR control with a smaller $m$ (see Appendix E).

In Figure \ref{fig:gsp}, we highlight the different choices of the global parameters $\xi'$s between the default horseshoe and the FAHS. Although both the default horseshoe and the FAHS preserve the large signals well, we learn the default horseshoe under-shrink the noise observations so that FDR cannot be guaranteed. The global parameters $\xi'$s in FAHS are verified to always satisfy Theorem 1 to achieve sharp Bayesian contraction rates.

\begin{figure}[t]
	\centering
	\includegraphics[width=0.9\linewidth]{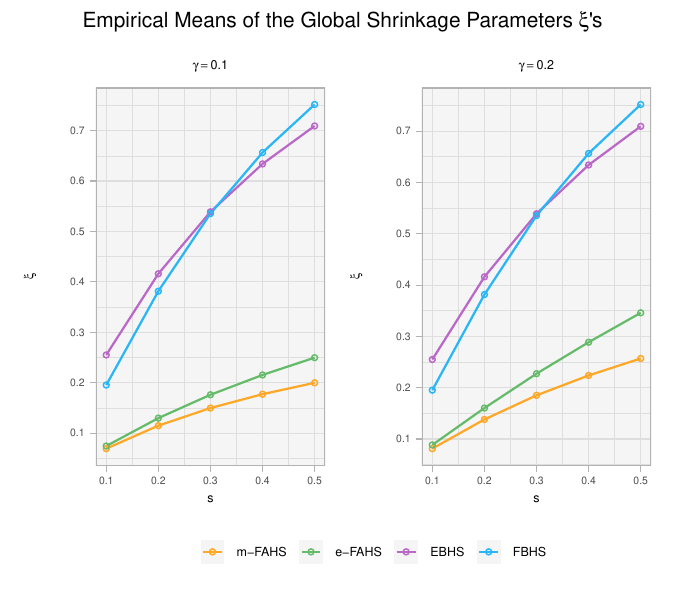}
	\caption{Empirical means of global shrinkage parameter $\xi$'s for four horseshoe procedures across 100 replications in the multiple independent testing setting.}
	\label{fig:gsp}
	\vspace{-10pt}
\end{figure}

\vspace{-3pt}
\subsection{Multiple Correlated Tests}

We set up the experiment in this section the same as the experiment in section \ref{sec4.1} except the vector $\boldsymbol{\epsilon}$ now has an equicorrelation structure with correlation $\rho$ being possibly 0.1, 0.2, and 0.3. The result with $\rho$ being 0.3 is shown in Figure \ref{fig_corr}. This is a case of famous PRDS condition \citep{Benjamini2001BY} and a case of $\mathrm{MTP}_2$ which is studied a lot in the graphical modeling literature \citep{wang2020GGMTP2}. The general pattern in this experiment is basically the same as the pattern of multiple independent tests in section \ref{sec4.1}.

\begin{figure}[t]
	\centering
	\includegraphics[width=\linewidth]{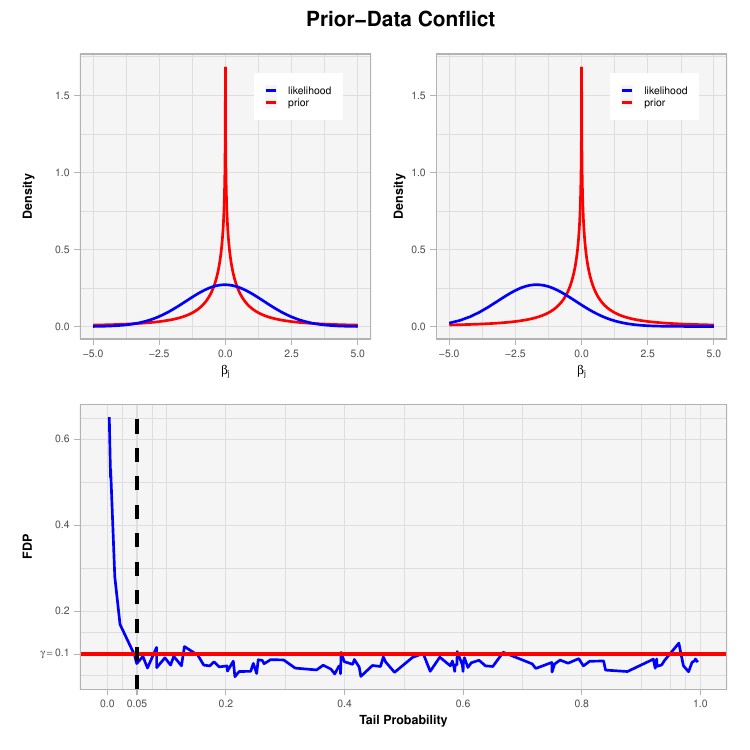}
	\caption{Top left: No prior-data conflict (i.e. horseshoe prior and data fit with each other) and the corresponding FDP is under control; Top right: Prior-data conflict exists and the corresponding FDP loses control greatly; Bottom: FDPs lose control more greatly when prior-data conflicts are more severe. The black dashed line is the pre-specified threshold for prior data conflicts. There is a noticeable increase for FDPs when the tail probabilities get lower and lower than the threshold.}
	\label{fig:pdc}
	\vspace{-0.5cm}
\end{figure}

The most notable difference between the two experiments is there are outliers (i.e. large FDPs) for every procedure except locfdr (but locfdr loses FDR control in most cases). This phenomenon is understandable with the involvement of correlation structure. However, as Bayesian, we can detect prior-data conflict (shown in Figure \ref{fig:pdc}) using a tail probability (e.g. a p-value) and replace the previous defective prior (horseshoe in our FAHS procedures) with a more suitable one using the weakly-informative prior methodology \citep{Gelman2006PD,Gelman2008AWI,evans2011wi} to reduce large FDPs. The individual FDPs are close to the FDR nominal level when we just begin to observe the prior-data conflict which provides us a reference to control overall FDR close to its nominal level. The algorithm for detecting prior-data conflict is in Appendix E.2.2.

Notice small $\gamma$ and s are often encountered in high-stakes scientific experiments \cite{Rudin2019}. We find that the empirical FDRs of both FAHS procedures are fairly close to the FDR nominal levels when the $\gamma$ and s are required to be small (see Figures from \ref{fig_ind_full} to \ref{fig_corr_0.3} for more details). Based on the tradeoff between FDR and power, our FAHS obtains reasonably high powers in those difficult cases.

\section{Real Data Study}
\label{sec5}

We demonstrate the practical application of m-FAHS and e-FAHS using a popular prostate cancer data set \citep{Singh2002GeneEC} which includes 6033 genes for 102 subjects, with 50 normal controls and 52 prostate cancer patients. The goal is to identify genes that are differentially expressed. To implement the analysis, we obtain $z$-scores for all 6033 genes. In this case, model (\ref{3normalmean}) is well suited here. Thus, this becomes the problem (\ref{4hypo}). By the strong and robust FDR control guarantee of FAHS, we discover two important new genes (i.e. gene 4331, gene 1113) compared with the well-established BH procedure. We list the top 10 important genes in Appendix E for BH, FAHS and locfdr procedures and recommend prioritizing studying genes on the order of the m-FAHS (which always gives the strongest FDR control) column for cost-effective reasons. More details and practical issues are discussed in Appendix E.2.3.

\section{Conclusions and Discussions}
\label{sec6}

With a happy marriage between frequentist and Bayesian, the FAHS procedures exhibit exceptional and encouraging performances. This suggests a new direction to design tests using the general class of global-local shrinkage priors: One can focus on the minimax construction and the idea of meff for the global parameter of such priors while keeping a suitable heavy-tail structure using the local parameter to retain the true signals. In our FAHS procedures, we assume the existence of p-values. There is an emerging methodology called e-values (\cite{vovk2021Evalue}; \cite{Ramdas2023CSEValue}) which has close connections with p-values but is more flexible and interpretable. It would be interesting to see what would happen for FAHS procedures if p-values were replaced by e-values.

\section{Acknowledgments}

The authors would like to thank Jinman Zhao for helpful discussions.

\nocite{langley00}

\bibliography{icml2025/mybib}
\bibliographystyle{icml2025}

\newpage
\appendix
\onecolumn

\section{Two-groups Model and One-group Model}
\label{secA}

\subsection{A brief comparison between the
two-groups model and the one-group
model}

We compare the two-groups model and the one-group model (e.g. horseshoe) in the high-dimensional normal means problem. Now, we have $m$ conditionally independent continuous observations $y=\left(y_1, y_2, \ldots, y_m\right)$ where $ y_j \mid \beta_j \sim N\left(\beta_j, 1\right).$ We want to test $H_{0j}: \beta_j=0, \ j=1, \ldots, m$. 

Given $s$, we suppose $\beta_j$'s are conditionally independent and the spike-and-slab type two-groups structure can be represented as 

$$
\beta_j \mid s \sim(1-s) \underbrace{\delta_{0}}_{\text {Spike }}+s \overbrace{N\left(0, \psi^2\right)}^{\text {Slab}}.
$$

Since the likelihood is given by $ y_j \mid \beta_j \sim N\left(\beta_j, 1\right),$ the marginal distribution of $y_j \mid s$ is a mixture of normals

$$
y_i \mid s \sim(1-s)N(0,1)+s N\left(0,1+\psi^2\right).
$$

The posterior mean $\mathbb{E}\left(\beta_j \mid y_j\right)$ under the two groups model is

$$
\mathbb{E}\left(\beta_j \mid y_j\right)=u_j \frac{\psi^2}{1+\psi^2} y_j=u_j^* y_j,
$$

where, $u_j$ is the posterior inclusion probability $P\left(\beta_j \neq 0 \mid y_j\right)$. We have $
\mathbb{E}\left(\beta_j \mid y_j\right) \approx u_j y_j
$ when $\psi^2 \rightarrow \infty$ as the number of tests $m \rightarrow \infty.$ We can see the $\psi^2$ term provides global shrinkage. The $u_j$ acts locally to adapt to the data's sparsity level.

Without using the discrete mixture, the one-group approach directly models the posterior inclusion probability $u_j^*$ instead of starting with a classification scheme to reach an estimator $u_j^* y_j$. \cite{carvalho09ahandingspHS} observed a suitable global-local mixture prior (i.e. a normal scale mixture in this case) leads to the same form of the posterior mean. The one-group model produces the analytically tractable marginal and behaves like a two-groups model using appropriate choices of the hyper-parameters. The normal scale mixture allows block-updating of the global and local shrinkage parameters which enables the fast computation for using the one-group model.

The horseshoe prior falls in a class of global-local shrinkage priors

$$
\begin{aligned}
y_j \mid \beta_j, \eta_j, \xi & \sim N\left(\beta_j, 1\right), \\
\beta_j \mid \eta_j, \xi & \sim N\left(0, \eta_j^2\right), \\
\eta_j \mid \xi & \sim C^{+}(0, \xi),
\end{aligned}
$$

where $C^{+}(0, \tau)$ is well known to be a heavy tailed prior. The posterior mean is $$\mathbb{E}\left(\beta_j \mid y_j, \xi\right)=\left(1-\mathbb{E}\left(\left.\frac{1}{1+\eta_j^2 \xi^2} \right\rvert\, y_j, \xi\right)\right) y_j,$$ where $\kappa_j=\frac{1}{1+\eta_j^2 \xi^2}.$ We find $1-\mathbb{E}\left(\kappa_j \mid y_j, \xi\right)$ mimics the posterior inclusion probability $P\left(\beta_j \neq 0 \mid y_j\right)$ by comparing the forms of posterior means for the two-groups model and the one-group model.

\begin{center}
\begin{tabular}{|c|c|}
\hline Two-groups Model & One-group Model \\
\hline$\mathbb{E}\left(\beta_j \mid y_j\right) \approx u_j y_j $ & $\mathbb{E}\left(\beta_j \mid y_j, \xi\right)=\left(1-\mathbb{E}\left(\kappa_j \mid y_j, \xi\right)\right) y_j$ \\
\hline
\end{tabular}
\end{center}

\subsection{Theoretical optimal properties of the horseshoe prior}

There are many amazing theoretical optimal properties for the horseshoe prior. Here, we list two of them as below.

1. \cite{Datta2013AsymHS} proved that the decision
rule induced by the horseshoe estimator is asymptotically Bayes
optimal for multiple testing under 0-1 loss.

2. Based on \cite{VanderPas2014HSconcen,VanderPas2017AdaptivePC}, they showed the horseshoe estimator has good
posterior concentration properties for nearly black objects. They also showed the horseshoe, horsehsoe+, and several other global-local estimators are minimax in $\ell_2$ up to a constant.

The half-Cauchy is a key component of horseshoe which has regularly-varying tails. Regular variation is closed under many nonlinear transformations which is a desirable property. Since the likelihood is light-tailed (i.e. normal in our case), the heavy-tailed prior helps in non-informative analysis \citep{Dawid1973MP}.

It is worth mentioning that \cite{Bhadra2017HSPlus} proposes a new prior called horseshoe+ which is a natural extension of the horseshoe with a better ability to separate signals from noises. The horsesheo+ has the hierarchical structure below,

$$
\begin{aligned}
y_j \mid \beta_j, \eta_j, \xi & \sim N\left(\beta_j, 1\right), \\
\beta_j \mid \eta_j, \xi & \sim N\left(0, \eta_j^2\right), \\
\eta_j \mid \xi & \sim C^{+}(0, \xi v_j), \\
v_j & \sim C^{+}(0, 1).
\end{aligned}
$$

Compared with the horseshoe estimator, the horsesheo+ estimator has a lower mean squared error, lower misclassification probability, and better posterior concentration properties in the Kullback-Liebler sense. It would be interesting and promising to apply the minimax or effective number of nonzero coefficients strategy to more flexible horseshoe+ to see the corresponding FDPs, FDRs, and stability.

\section{Remarks about FDR Control Procedures }
\label{secB}

We will use the notations in the following table throughout Section \ref{secB}.

$$
\begin{array}{cccc}
\hline &  H_0 \text{ Retained } &  H_0 \text { Rejected }& \text { Total } \\
H_0 \text { True (Noise)} & \text{TN} & \text{FD} & m_0 \\
H_0 \text { False (Signal)} & \text{FN} & \text{TD} & m_1 \\
\text { Total } & m-\text{R} & \text{R} & m=m_0+m_1 \\
\hline \hline
\end{array}
$$

\subsection{Remarks about the BH procedure}

\cite{Knight2024mht} gives a simple and intuitive heuristic explanation of why the Benjamini-Hochberg (BH) procedure works. Suppose that $m$ is large and assume that for some $\epsilon'>0$, approximately $(1-\epsilon') m$ null hypotheses are true and approximately $\epsilon' m$ are false. Let $\mathcal{S}=\left\{l: H_{0l}\right. \text{is true}\}$ and we assume that the p-values $\left\{p_l: l \in \mathcal{S}\right\}$ are uniformly distributed on $[0,1]$ while $\left\{p_l: l \notin \mathcal{S}\right\}$ have a distribution function $G$ that is stochastically smaller than the uniform distribution. Then, the distribution function of the p-values is given by $H(x)=\epsilon' G(y)+(1-\epsilon') y$. Now for some $0<\gamma<1$, we define

$$
\psi=\sup \{t \geq 0: \epsilon' G(\gamma t)+(1-\epsilon') \gamma t \geq t\}=\sup \left\{t \geq 0: H^{-1}(t) \leq \gamma t\right\}.
$$

Notice that $\psi$ may equal 0 but we will assume that $\psi>0$. In the BH procedure, we can think of $\hat{l} / m$ as an estimator of $\psi$ with $p_{(l)}$ an estimator of the quantile $H^{-1}(l / m)$, which represents the proportion of rejected null hypotheses. Then

$$
\text{FDP} = \frac{\text{FD}}{\text{R}}=\frac{\text { number of false rejections }}{\text { total number of rejections }} \approx \frac{(1-\epsilon') \gamma \psi}{\psi} \leq \gamma \text {. }
$$

When $\epsilon'$ is small (i.e. almost all null hypotheses are true), $\text{FD} / \text{R}$ will be close to $\gamma$ when $m$ is large. We also provide a graphical illustration (i.e. Figure 6) of the BH procedure (using notations in Algorithm 1 of subsection 2.2) below. 

\begin{figure*}[ht] 
    \centering
    \includegraphics[width=0.6\textwidth]{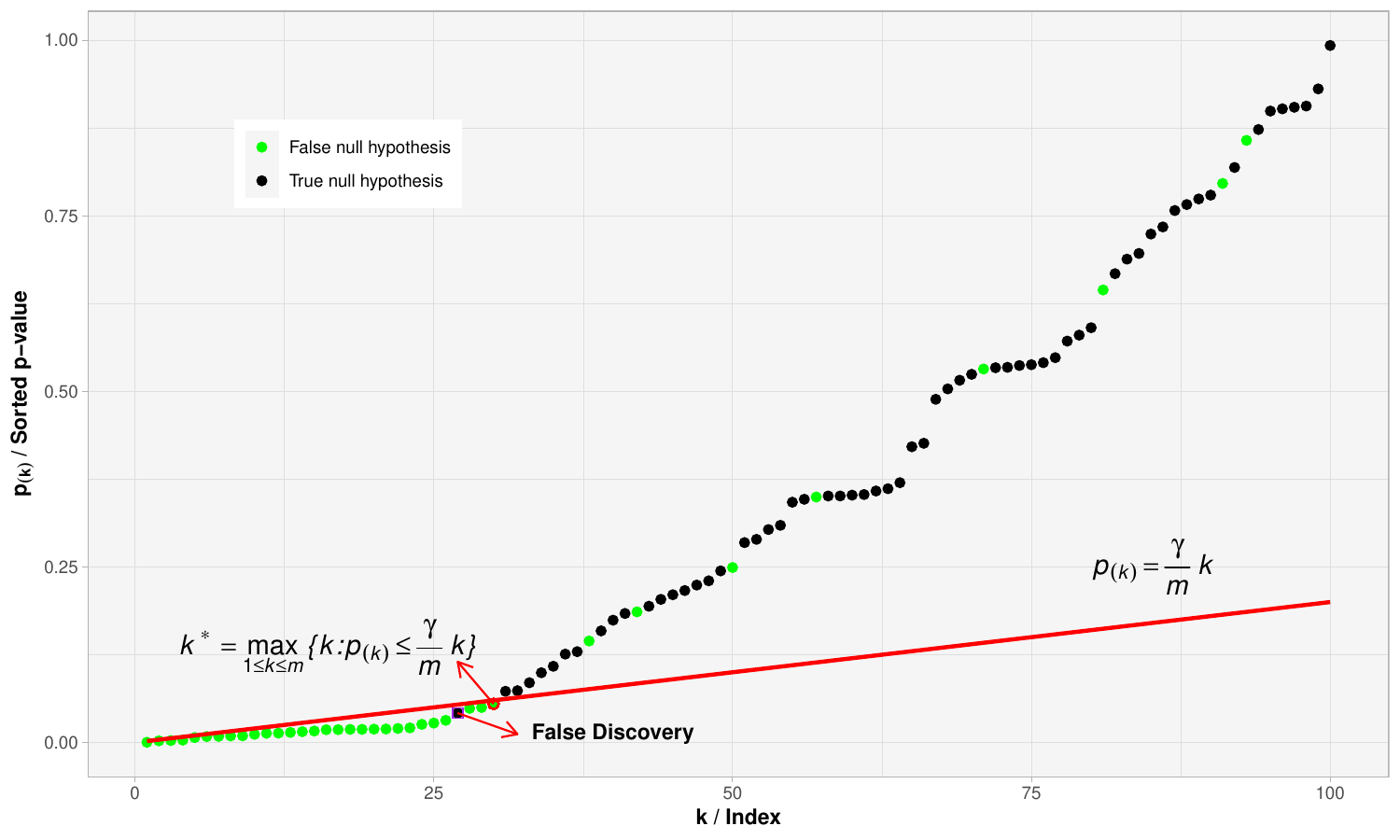} 
    \caption{Illustration of the Benjamini-Hochberg procedure controlling the false discovery rate.}
    \label{fig_bh}
\end{figure*}

\subsection{Remarks about the q-value procedure}
Although the q-value has certain mathematical equivalences with the p-value, they do have differences in terms of interpretations. Notice that adjusted p-values are defined in terms of a particular test procedure, while q-values are not. Thus, one should be careful about p-values and q-values when the interpretations of them are needed. Another note is that the Bayesian interpretation of the q-value is not classical Bayesian because the conditioning is on a set representing tail behavior rather than on the data. 

The q-value is also closely connected with the positive false discovery rate (pFDR). Although the pFDR has many similarities with the FDR, there are surprisingly different properties between the pFDR and FDR. Section 11.5 of \cite{Clarke2009DMML} is devoted to this.

\subsection{Remarks about the two-groups empirical Bayes procedure}

In terms of the algorithm, the two-groups empirical Bayes algorithm (i.e. Algorithm 2 in subsection 2.2) has a similar flavor to the BH algorithm (i.e. Algorithm 1 in subsection 2.2). In Algorithm 2, the estimated locfdr replaces the position of the p-value in Algorithm 1. The threshold for Algorithm 2 is determined by a moving average instead of finding the largest p-value that falls below the diagonal cutoff line in Algorithm 1. There are rich connections among the BH,  the q-value, and the two-groups empirical Bayes procedures from different perspectives, interested readers can refer to \cite{Efron2010LSI} and \cite{Clarke2009DMML}.

\cite{Efron2010LSI} uses Poisson regression with splines to
model histogram counts while the implementation can
be highly unstable. \cite{Sun2007OracleAA} use the kernel
density estimator to do the density estimation with the
bandwidth chosen by cross-validation. The bandwidth
is sensitive and arguably most crucial for a kernel density estimator but cross-validation is not a cure-all.
The computation of cross-validation is usually quite
heavy when there are many observations. When observations are only a few, the result of cross-validation
can be misleading. In terms of the kernel density estimator itself, it is known to easily break down in the
high dimension so it does not scale up well. There
have been many more proposals for the deconvolution
problem since then, but the majority of them are either overly complicated to use or use oversimplified
and unrealistic assumptions.

Some want to save resources not to estimate $f_0$. Instead, they use the theoretical null distribution for $f_0$. However,  $z$-values should follow their theoretical null distributions under the null when all the assumptions are true. It is certainly possible, for a variety of reasons, for the theoretical nulls not to hold so that theoretical nulls are quite different from the empirical nulls in applications.

No matter the two-groups empirical Bayes procedure is developed to be model-free or not, virtually all those kinds of procedures only asymptotically control FDR. Based on the experiments in this paper, we find the two-groups empirical Bayes procedure strictly controls FDR under the nominal level in practice only when the signals are relatively dense, the FDR nominal level is relatively high, or both.

\subsection{Remarks about the vanilla horseshoe}
It is worth noting that \cite{scott2010BM} beautifully handle the Bayesian multiplicity issue in the variable selection context for the Bayesian hierarchical model by assigning a suitable prior structure. However, their strategy cannot be trivially transformed to tackle the FDR control problem using the horseshoe prior or general class of global-local shrinkage priors. Thus, we have to figure out other ways to design tests with FDR control using the horseshoe prior or the general global-local shrinkage prior family.

\section{Technical Definitions, Assumptions and Explanations Related to Theorem \ref{theo1} in Section \ref{sec3}}
\label{secC}

We first define two notations $\succ$ and $\asymp$ here. Given two positive sequences $\left\{a_n\right\}$ and $\left\{b_n\right\},$ we define $a_n \succ b_n$ so that $\lim \left(a_n / b_n\right)=\infty.$ We also define $a_n \asymp b_n$ which means $0<\liminf \left(a_n / b_n\right) \leq \lim \sup \left(a_n / b_n\right)<\infty.$

Theorem 1 in Section 3 (Theorem 2.1. of \cite{song2020BSSM}) studies the relationship between the polynomial order of $\pi_0$ and the posterior contraction rate. It also establishes the Bayesian sharp minimax. We require three assumptions on the model sparsity and prior distribution $\pi_0$ (which has a polynomial tail) for Theorem 1 to hold in the sparse normal means context.

\textbf{C.1.1} The true model is sparse.

\textbf{C.1.2} The prior density $\pi_0(\cdot)$ is strictly decreasing on $(0, \infty)$ and increasing on $(-\infty, 0)$.

\textbf{C.1.3} The tail of $\pi_0(\cdot)$ is polynomially decaying with polynomial order $\alpha>1.$ In other words, there exist some positive constants $L$ and $C_2>C_1$ such that $C_1|\boldsymbol{\beta}|^{-\alpha} \leq \pi_0(\boldsymbol{\beta}) \leq C_2|\boldsymbol{\beta}|^{-\alpha}$ for any $|\boldsymbol{\beta}|>L$.

It is clear that the horseshoe prior satisfies these three assumptions in the sparse normal means problem. Notice that \textbf{C.1.3} implies the polynomial decaying of $\pi\left(\beta_j \mid \xi\right)$. In other words, if $\left|\beta_j\right| \gg \xi$, then $\pi\left(\beta_j \mid \xi\right) \asymp\left|\beta_j\right|^{-\alpha} \xi^{\alpha-1}$.

Bayesian variable selection used to be implemented via the spike-and-slab priors. As we introduced in Section 1.1, it has a two-group structure, i.e.,
\[
\beta_j \sim (1-s)\delta_0 + sg(\beta_j),j=1,\cdots,m
\]
where $\delta_0$ is a Dirac point mass at 0 used to model noise, $g(\beta_j)$ is an absolute continuous density used to model signals and $s$ represents the proportion of signals. Due to the computational difficulty of $\delta_0$, in practice, researchers tend to use an absolute continuous density that concentrates around 0 to replace $\delta_0$. However, the binary mixture splits the parameter space into $2^m$ spaces, making it difficult for Markov Chain Monte Carlo (MCMC) algorithms to explore the entire parameter space. Hence, in high-dimensional problems, the spike-and-slab priors become increasingly inefficient as the complexity grows exponentially. To overcome this issue, \cite{Carvalho2010TheHE} introduced the global-local shrinkage priors. After that, the horseshoe prior has become one of the most famous global-local shrinkage priors because it is easy to implement and has comparable performance over traditional statistical problems. 

Although \cite{Carvalho2010TheHE} proposed the horseshoe prior, they did not discuss the optimal choice of global shrinkage parameter $\xi$. To complete the discussion, researchers need to find one criterion. Note that Bayesian decision theory carries out inference problems (estimation, testing, etc.) using different loss functions. The researchers then evaluate and minimize the posterior risk functions which are the expectations of loss functions over the posterior distribution. Hence, for different inference tasks, the common atom is the posterior distribution. If the posterior distribution is concentrated around the ground truth, any inference problem could be solved easily. That is why researchers would like to choose $\xi$ such that the posterior distribution is the most concentrated around the ground truth. The main tool is the posterior concentration inequality
$$
\lim _{m \rightarrow \infty} \mathbb{E}^*\left[\pi\left(\left\|\boldsymbol{\beta}-\boldsymbol{\beta}^*\right\| \geq r_m \mid \boldsymbol{y}\right)\right]=0, \text { for any } \boldsymbol{\beta}^*,
$$
where $\boldsymbol{\beta}^*$ is the truth, $\pi(\cdot|\boldsymbol{y})$ is the posterior density, $r_m$ is the concentration rate and $E^*$ represents the expectation taken over $g(\boldsymbol{y}|\boldsymbol{\beta}^*)$. Researchers usually set the target $r_m$ as the minimax rate in \cite{Donoho1992ME}, i.e., $r_m^2 = (2+o(1))m_1\log(m/m_1)$, and try to establish the conditions of $\xi$ to ensure the inequality holds. There are several papers discussing the choice of $\xi$ through the idea of posterior concentration. We included \cite{VanderPas2014HSconcen} and \cite{song2020BSSM} in our paper because the former is the first to conduct such a discussion, while the latter provides the most general results. \cite{VanderPas2014HSconcen} established their results by Markov inequality and proposed the conditions of $\xi$, i.e., $\xi = (m_1/m)^{\alpha},\alpha \ge 1$. Combining with the results of \cite{Datta2013AsymHS}, we consider $\xi^*=m_1/m$ in our paper. \cite{song2020BSSM} provided a better interpretation on how the tail order of prior density relates to the conditions of $\xi$. For example, we use the horseshoe prior in our paper, whose tail is polynomially decaying of order 2. Hence, in Theorem 1 of our paper, $\alpha = 2$. According to Corollary 2.1 of \cite{song2020BSSM}, the constant $w$ should satisfy $w \ge 2(\alpha-1)=2$. We hope the concentration rate to be the smallest, so we choose $w = 2$. As a result, in Theorem 1 in our paper, the condition of $\xi$ should be
\[
(m_1/m)^c \sqrt{\log(m/m_1)} \le \xi \prec ((m_1/m)\log(m/m_1))^2,
\]
where $c \in (0,1)$. Any $\xi$ in this interval guarantees the target concentration rate. This result helps us validate our choices of $\xi$ based on \cite{VanderPas2014HSconcen} and \cite{Piironen2017SparsityIA}. Note that \cite{Piironen2017SparsityIA} choose $\xi$ by letting $\mathbb{E}[\sum_{j=1}^m \kappa_j|\boldsymbol{y}]=m_1$, which establishes a non-asymptotic result. Meanwhile, unifying the results of \cite{VanderPas2014HSconcen} and \cite{Piironen2017SparsityIA} into the theory of \cite{song2020BSSM} also benefits from the theoretical results linking posterior contraction rate to FDR control established by \cite{song2023FDRMm}. It provides the theoretical guarantee of our procedures but lacks of empirical evidence, which is now filled by this paper.

\section{Further Study of Global Parameter of the Horseshoe}
\label{secD}

The following Figure 7 shows the importance of setting the global parameter. Both default FBHS and EBHS can not obtain enough sparsity around the original and then lose control of FDR. In comparison, both FAHS procedures can acquire sufficient sparsity and preserve large signals using the heavy tail to control finite-sample FDR consistently. Notice that the yellow line (i.e. m-FAHS) is close to the green line (i.e. e-FAHS).

\begin{figure*}[ht] 
	\centering
	\includegraphics[width=0.5\textwidth]{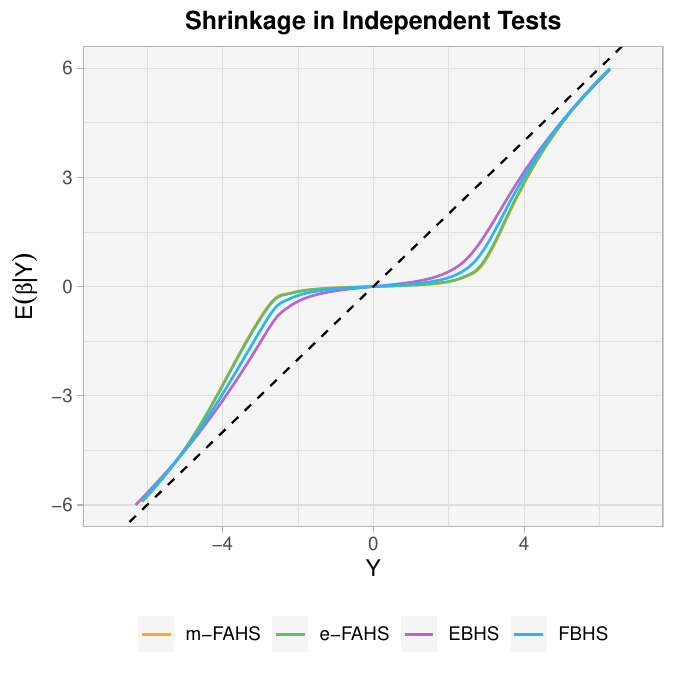} 
	\caption{Illustration of shrinkage for the horseshoe methods (m-FAHS, e-FAHS, EBHS, and FBHS) in the multiple independent testing setting, where m-FAHS and e-FAHS obtain more sparsity near the original and still preserve large signals. In this experiment, we test 10,000 independent hypotheses simultaneously under random seed 1 with signal proportion being 0.05 and FDR nominal level being 0.1.}
	\label{fig_sh}
\end{figure*}

\section{Comprehensive Plots and Details for Section 4 and Section 5}
\label{secE}

\subsection{Data generating process and implementation details of related horseshoe procedures}

\subsubsection{Data generating process}
We use the following three steps to generate the sparse normal means as the underlying truth in Section 4.

Step 1. Generate $\boldsymbol{\theta} \sim$ rbinom $(m, 1, s)$.

Step 2. Generate $\boldsymbol{\beta}=\boldsymbol{\theta} \cdot \operatorname{rnorm}(m, 0, \psi^2),$ then standardize $\boldsymbol{\beta}$ and multiply $\boldsymbol{\beta}$ by the signal-to-noise ratio (SNR) (SNR is 3 here).

Step 3. Generate $\boldsymbol{y}=\boldsymbol{\beta}+\operatorname{rnorm}(m, 0,1).$

\subsubsection{Implementation details of related horseshoe procedures}
The implementation details of related horseshoe procedures used in the paper are summarized in the following Table 2. 

\medskip

\begin{table}[ht]
    \centering
    \begin{threeparttable}
        \caption{Implementation details of horseshoe methods.}
        \begin{tabular}{cccc}
        \toprule
             & Global Shrinkage Parameter $\xi$ & Local Shrinkage Parameter $\eta_j$'s & Model Variance $\sigma^2$ \\
             \midrule
             \textbf{m-FAHS} & $\frac{\text{R}_{BH}}{m}$ & standard half-Cauchy prior & Jeffreys prior\\
             \addlinespace
             \textbf{e-FAHS} & $\frac{\sigma \text{R}_{BH}}{m-R_{BH}}$ & standard half-Cauchy prior & fixed at 1\\
             \addlinespace
             \textbf{EBHS} & estimated by MMLE & standard half-Cauchy prior & Jeffreys prior\\
             \addlinespace
             \textbf{FBHS}& standard half-Cauchy prior & standard half-Cauchy prior & Jeffreys prior\\
        \bottomrule
        \end{tabular}

        \begin{tablenotes}
            \footnotesize
            \item[*] $\text{R}_{BH}$ refers to the number of total discoveries obtained by the Benjamini-Hochberg procedure.
        \end{tablenotes}
    \end{threeparttable}
    \label{tab1}
\end{table}

\subsection{Comprehensive plots for FDR control
procedures and the prior-data conflict algorithm in Section 4}

\subsubsection{Comprehensive comparisons among FDR control procedures}

\begin{figure*}[hb] 
	\centering
	\includegraphics[width=\textwidth, scale =0.1]{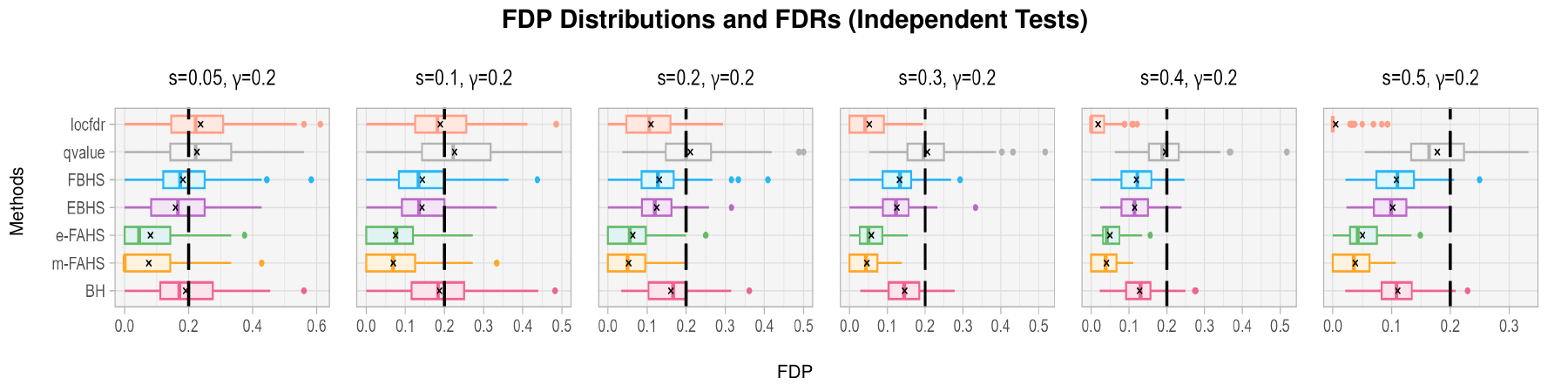} 
	\caption{FDP distributions (boxplots) and FDRs (black crosses) for multiple FDR control methods in the multiple independent testing setting. Black long dashed lines refer to nominal FDR levels. In this experiment, we do 100 replications where we test 200 independent hypotheses simultaneously in each replication under the signal proportion ranging from 0.05 to 0.5 and the nominal FDR level being 0.2.}
	\label{m=200}
\end{figure*}

\begin{figure*}[ht] 
	\centering
	\includegraphics[width=\textwidth]{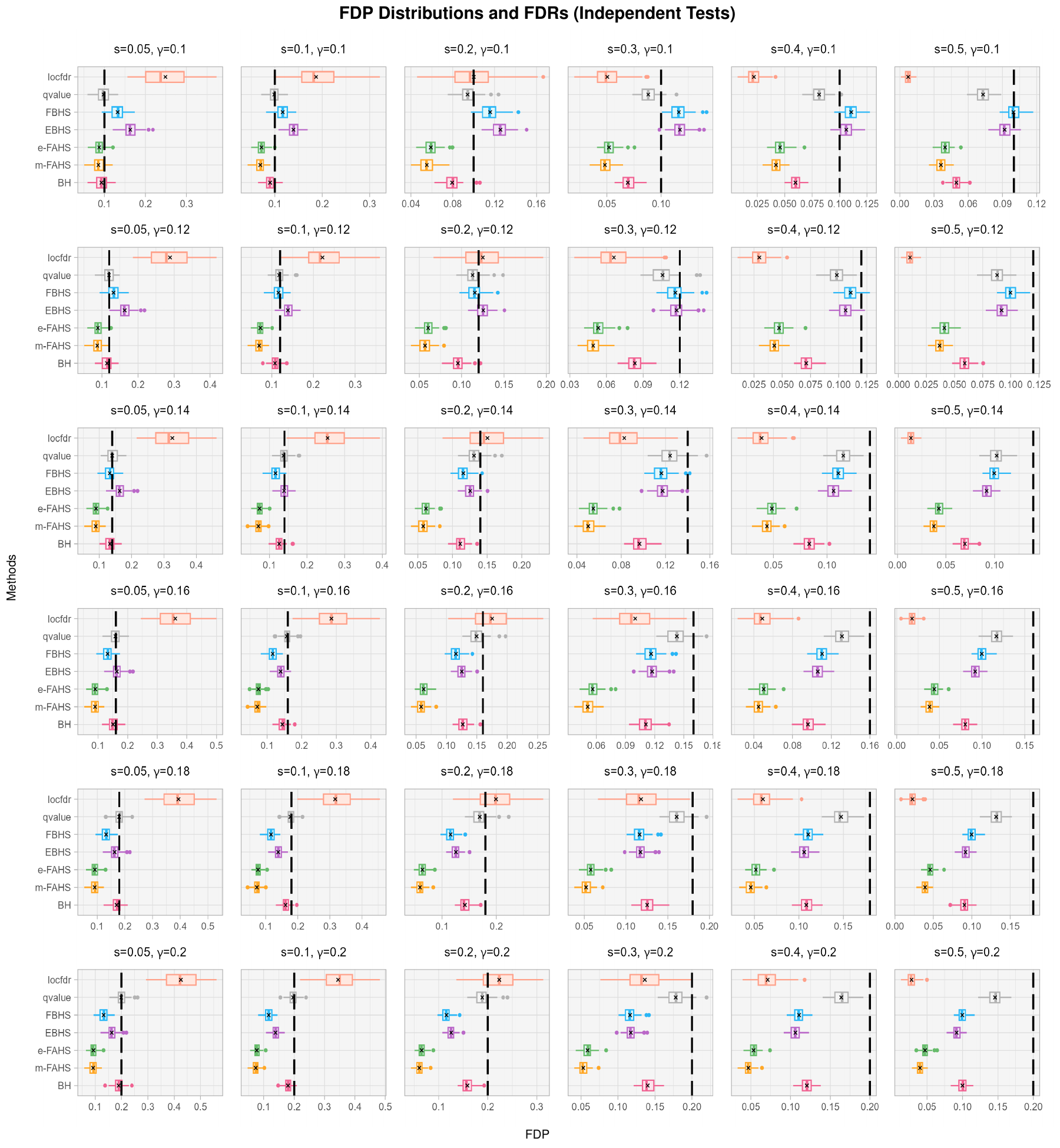} 
	\caption{FDP distributions (boxplots) and FDRs (black crosses) for multiple FDR control procedures in the multiple independent testing setting where m-FAHS and e-FAHS show more robust FDR and FDP controls than any other procedure. Black vertical dashed lines refer to FDR nominal levels. In this experiment, we do 100 replications where we test 10000 independent hypotheses simultaneously in each replication under the signal proportion ranging from 0.05 to 0.5 and the FDR nominal level ranging from 0.1 to 0.2.}
	\label{fig_ind_full}
\end{figure*}

\begin{figure*}[ht] 
	\centering
	\includegraphics[width=\textwidth]{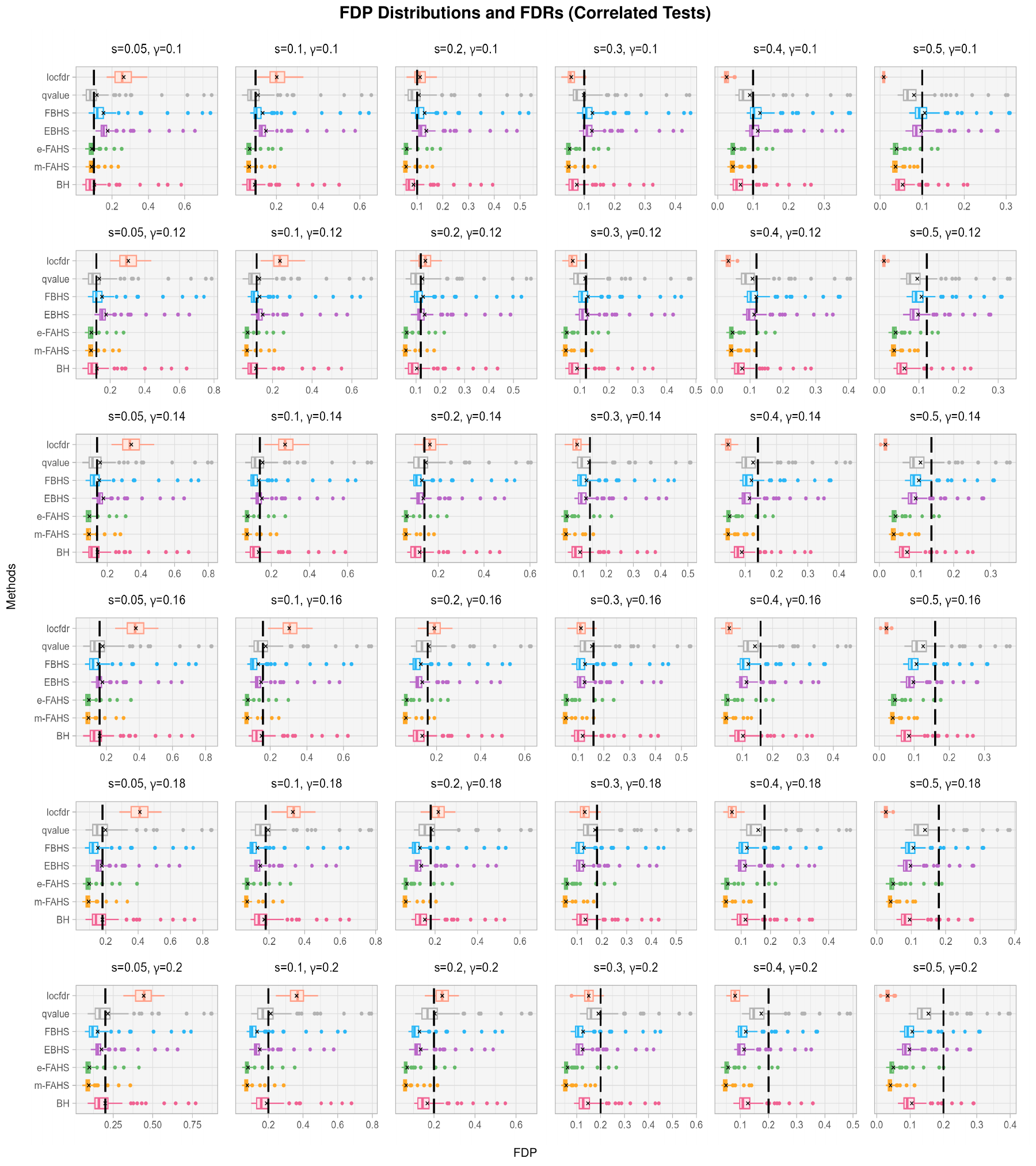} 
	\caption{FDP distributions (boxplots) and FDRs (black crosses) for multiple FDR control methods in the multiple correlated testing setting (equicorrelation structure with equal correlation 0.1) where m-FAHS and e-FAHS show more robust FDR and FDP controls than any other procedures. Black vertical dashed lines refer to nominal FDR levels. In this experiment, we do 100 replications where we test 10000 correlated hypotheses simultaneously in each replication under the signal proportion ranging from 0.05 to 0.5 and the nominal FDR level ranging from 0.1 to 0.2.}
	\label{fig_corr_0.1}
\end{figure*}

\begin{figure*}[ht] 
	\centering
	\includegraphics[width=\textwidth]{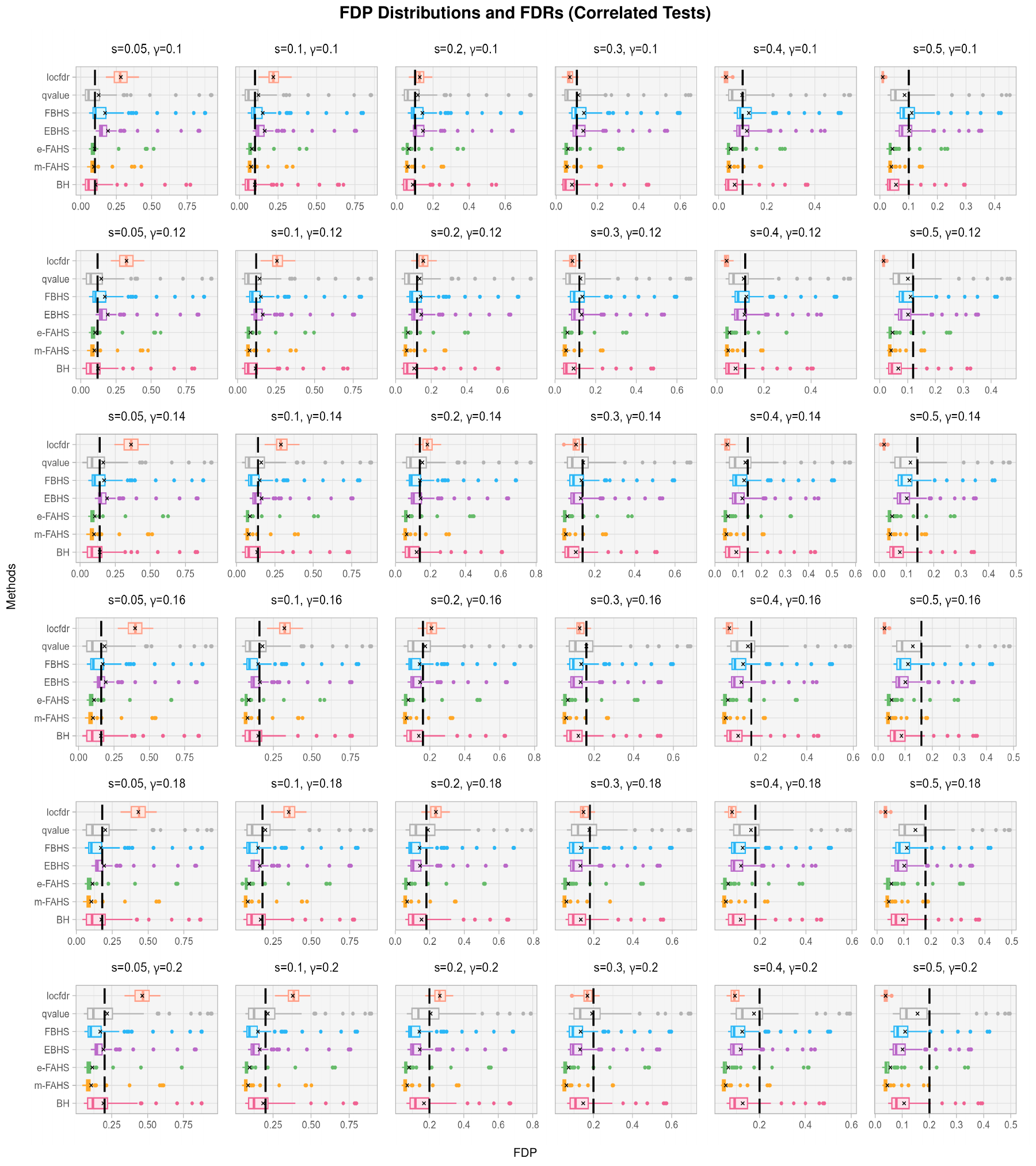} 
	\caption{FDP distributions (boxplots) and FDRs (black crosses) for multiple FDR control methods in the multiple correlated testing setting (equicorrelation structure with equal correlation 0.2) where m-FAHS and e-FAHS show more robust FDR and FDP controls than any other procedures. Black vertical dashed lines refer to nominal FDR levels. In this experiment, we do 100 replications where we test 10000 correlated hypotheses simultaneously in each replication under the signal proportion ranging from 0.05 to 0.5 and the nominal FDR level ranging from 0.1 to 0.2.}
	\label{fig_corr_0.2}
\end{figure*}

\begin{figure*}[ht] 
	\centering
	\includegraphics[width=\textwidth]{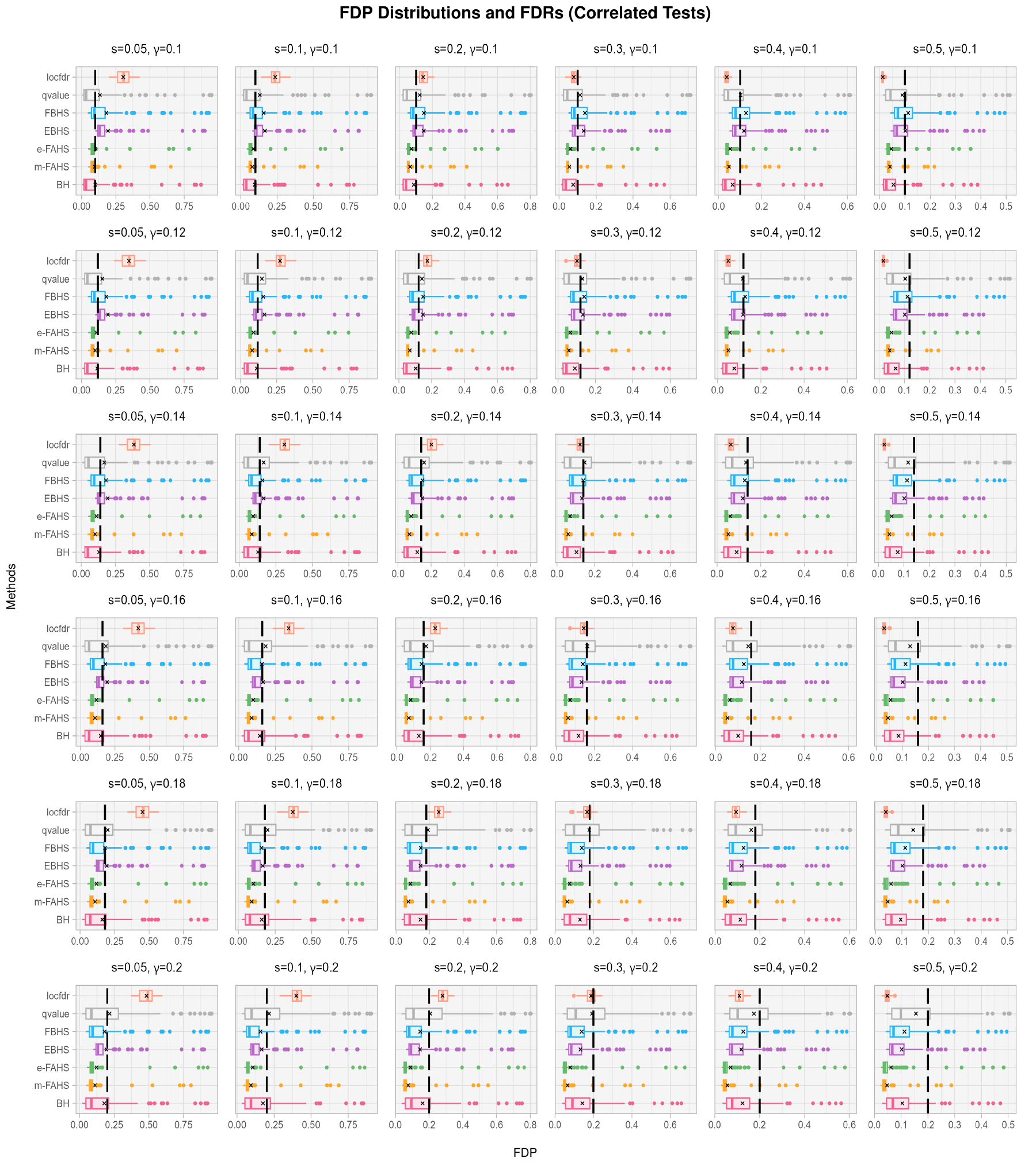} 
	\caption{FDP distributions (boxplots) and FDRs (black crosses) for multiple FDR control methods in the multiple correlated testing setting (equicorrelation structure with equal correlation 0.3) where m-FAHS and e-FAHS show more robust FDR and FDP controls than any other procedures. Black vertical dashed lines refer to nominal FDR levels. In this experiment, we do 100 replications where we test 10000 correlated hypotheses simultaneously in each replication under the signal proportion ranging from 0.05 to 0.5 and the nominal FDR level ranging from 0.1 to 0.2.}
	\label{fig_corr_0.3}
\end{figure*}

Through Figures from \ref{m=200} to \ref{fig_corr_0.3}, we show comprehensive comparisons among a variety of FDR control procedures we discuss in the paper. No matter there is correlations among tests or not, our m-FAHS and e-FAHS always achieve finite-sample FDR control and show strong robustness in any setting. In the main text, we have illustrated the performance of locfdr and vanilla horseshoe in great length. Here, we want to particularly point out there are cases the q-value procedure loses FDR control in the multiple correlated testing setting (Figure \ref{fig_ind_full} to Figure \ref{fig_corr_0.3}). We do an extra experiment in the multiple independent testing setting (Figure \ref{m=200}), q-value loses FDR control in several sparse scenarios (i.e. s=0.05, 0.1, and 0.2) when the FDR nominal level is relatively high (e.g. $\gamma=0.2$) and the number of hypotheses is relatively small (e.g. 200).

\bigskip

\clearpage
\subsubsection{Prior-data conflict algorithm in subsection 4.2}

 In the multiple correlated testing setting, there are some large FDPs for most procedures. This is mainly due to the prior-data conflict which refers to prior may place most of its mass on parameter values that are not feasible based on the data. In the main text, we briefly illustrate how we can detect them using the idea of prior-data conflict.  The concrete prior-data conflict algorithm is given as Algorithm 4.
 
\begin{algorithm}[ht]
\caption{The prior-data conflict algorithm for horseshoe prior}
\begin{algorithmic}[1]

\State For each random seed we sample simulated data and sample the local shrinkage parameters $\eta_j, j=1,2,\dots,10000.$ from a half-Cauchy distribution independently.

\State Take the mean of $y_j$'s, (i.e. the sufficient statistic $\bar{{y}}$) and calculate the prior predictive distribution of $\bar{{y}}$ in the equal-correlated normal means model, i.e. $$N\left(0,\frac{\sigma_{jj}^2 \xi^2 \sum_{j=1}^m{\eta_j^2}}{m^2}+\frac{\sigma_{jj}^2+(m-1)\sigma_{jk}^2}{m}\right),$$ where $\xi$'s are plugged in by estimated global shrinkage parameters in m-FAHS and e-FAHS, $\sigma_{jj}$'s and $\sigma_{jk}$'s are the true diagonal and off-diagonal entries in the covariance matrix $\Sigma$ of $\boldsymbol{y}$ in our settings.

\State Calculate the closed-form prior-data conflict tail probability for some particular data generating process, i.e.  
$$
P \left(|\bar{{Y}}| < |\bar{{y}}| \right) = 2 \left(1-\boldsymbol{\Phi} \left(\frac{|\bar{{y}}|}{\sqrt{\frac{\sigma_{jj}^2 \xi^2 \sum_{j=1}^m{\eta_j^2}}{m^2}+\frac{\sigma_{jj}^2+(m-1)\sigma_{jk}^2}{m}}}\right)\right),
$$
where $\boldsymbol{\Phi}$ refers to the cumulative distribution function of the standard normal distribution.

\State Reject that there is no prior-data conflict in the seed if the tail probability is less than the threshold, say 0.05.

\end{algorithmic}
\end{algorithm}

\subsubsection{More about the real data}
The dataset we use consists of expression levels for $6033$ genes from 52 cancer patients and 50 subjects in the control group. Let $z_{ji}$ be the expression level for gene $j$ on subject $i$ for $j=1,\cdots,6033$ and $i=1,\cdots,102$. The subjects in the control group are indicated by subscripts $i=1,\cdots,50$ and the cancer patients by $i=51,\cdots,102$. The $t$-statistic for comparing the gene expression levels between the two groups is computed for each gene as $t_j=\frac{\bar z_{2j}-\bar z_{1j}}{s_j}$ where $\bar z_{1j}=\frac{1}{50}\sum_{i=1}^{50}z_{ji}$, $\bar z_{2j}=\frac{i}{52}\sum_{i=51}^{102}z_{ji}$ and  
$$s_j^2=\left( \frac{1}{50}+\frac{1}{52} \right) \frac{\sum_{i=1}^{50}(z_{ji}-\bar z_{1j})^2+\sum_{i=51}^{102}(z_{ji}-\bar z_{2j})^2}{100}.$$ Let $\Phi$ be the cumulative distribution function (CDF) of the standard normal distribution and $F_{t_{100}}$ be the CDF of the $t$ distribution of 100 degrees of freedom. According to \cite{Efron2010LSI} Chapter 2, we use $x_i=\Phi^{-1}(F_{t_{100}}(t_j)), j = 1, \ldots, m$, as the observed data. 

Under the global null hypothesis (i.e. no differentially expressed genes), one would expect the histogram of the test statistics to follow a $N(0,1)$ density curve. However, we find the histogram (i.e. Figure 13) shows a heavier tail which suggests the existence of some differentially expressed genes (i.e. signals).

\begin{figure*}[htbp] 
	\centering
	\includegraphics[width=0.9\textwidth]{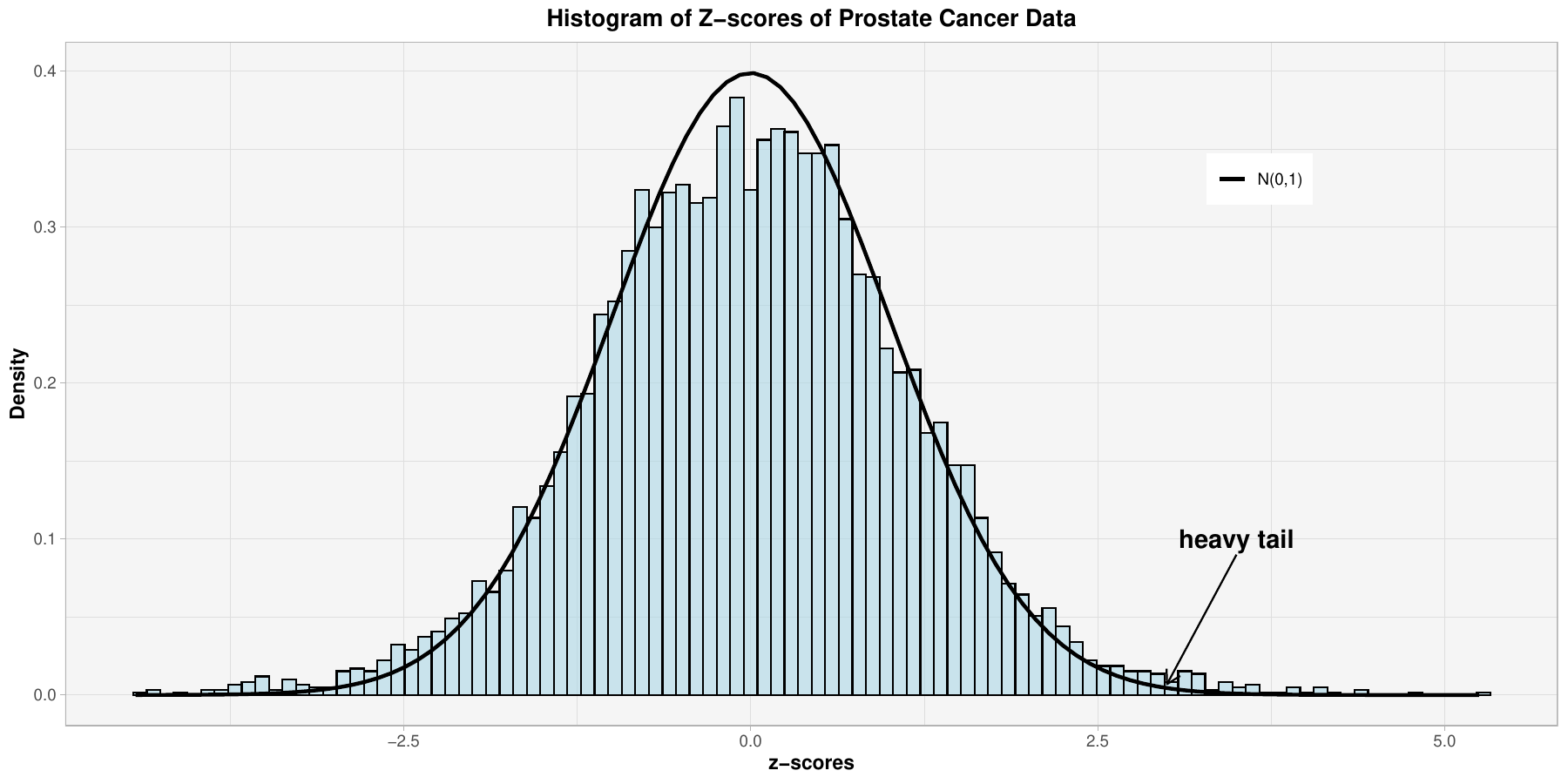} 
	\caption{Z-scores (histogram) for the prostate cancer data, where the heavy tail indicates that there are some signals in the 6,033 z-scores.}
	\label{fig_pdc_e}
\end{figure*}

We list the top 10 important genes in Table 3 for BH, FAHS and locfdr
procedures and recommend prioritizing studying genes
on the order of the m-FAHS (which always gives the
strongest FDR control) column for cost-effective reasons. The gene numbers with the underscore are genes found important by FAHS procedures but not by the BH procedure.

The original bulk RNA sequencing data is not available due to confidentiality. We downloaded the dataset from \cite{Efron2010LSI}, which only contains the expression levels of genes in each sample without the ID of genes. Hence, it is difficult to speak for the 2 more genes we discovered by investigating the existing literature on pathway study. However, this dataset has been widely used since \cite{Efron2010LSI}, such as \cite{Carvalho2010TheHE}, \cite{Bhadra2017HSPlus} and \cite{Ghosal2017NBI}. When the Bayes risk or FDR can be controlled according to theory or simulation study, the more signals found in this real dataset the better. Based on this idea, we believe the 2 new genes detected by our procedure could be positive evidence of the empirical performance of our method. 

\begin{table}[ht]
\caption{Top 10 discovered genes in the prostate cancer data (where $\gamma$ = 0.1)
} 
\label{table_3}
\begin{center}
\begin{tabular}{rrrr}
\toprule
\textbf{BH} & \textbf{m-FAHS} & \textbf{e-FAHS} & \textbf{locfdr}\\
\midrule
610 & 610 & 610 & 610\\
1720 & 1720 & 1720 & 1720\\
332 & 364 & 332 & 364\\
364 & 914 & 364 & 3940\\
914 & 3940 & 4546 & 4546\\
3940 & 332 & 914 & 332\\
4546 & \underline{\textbf{4331}} & 3940 & 914\\
579 & 4546 & \underline{\textbf{4331}} & 4331\\
1068 & 1068 & 1068 & 1068\\
1089 & \underline{\textbf{1113}} & 1089 & 579\\
\bottomrule
\end{tabular}
\end{center}
\end{table}

\section{Potential Extensions of Current Work}
\label{secF}

Since m-FAHS and e-FAHS procedures are based on the theory of posterior contraction rates and the effective number of signals from \cite{Piironen2017SparsityIA}, respectively, we would like to expand our discussion in two aspects. For m-FAHS, since the theory of posterior contraction in sparse generalized linear models (GLM) has been established by \cite{Jeong2021PC}, there is no theoretical gap to extend m-FAHS into GLM to accommodate other data types such as binary data, categorical data and count data. In addition, although the discussion in \cite{VanderPas2014HSconcen} was limited respect to the horseshoe prior, the more general results provided by \cite{song2020BSSM} only assumed the polynomial decaying order of the marginal prior on the parameter of interest. Therefore, m-FAHS can naturally be applied to hierarchical models as long as the tail of the marginal prior is polynomially decaying. For e-FAHS, \cite{Piironen2017SparsityIA} already provided the calculation of effective number of signals in GLM via Laplace approximation of the likelihood. Hence, there is also no theoretical gap to extend e-FAHS to more complex models. Their calculation covered the horseshoe prior and the regularized horseshoe prior. The latter would give us the chance to also extend e-FAHS to statistical problems with more dense signals. To summarize, there is no gap to extend m-FAHS and e-FAHS to GLM, as well as m-FAHS to other global-local shrinkage priors, but as \cite{Piironen2017SparsityIA} mentioned in their paper, when the models get more complicated like Bayes trend filter of \cite{Faulkner2017LAS} and horseshoe Bayesian neural network of \cite{Ghosal2017NBI}, the interaction between global parameter $\xi$ and the effect of sparsity remains an unsolved problem. However, we believe as the theory improves, m-FAHS and e-FAHS can easily keep up with it. We also want to point out here one may replace BH with some
other adaptive frequentist FDR control procedures such as e-values (which have
a close relation with another popular knockoff approach \citep{ren2024e})
mentioned in our conclusion section. E-values can be interpreted as Bayes fac-
tors \citep{Ramdas2023CSEValue} which are interesting in themselves. As a future goal,
we hope we can gradually develop FAHS to be a potential framework to bridge
the frequentism and Bayesian in an empirical Bayes manner.


\balance
\end{document}